\documentclass[12pt]{article}
\usepackage{amsmath}
\usepackage{graphicx,psfrag,epsf}
\usepackage{enumerate}
\usepackage{natbib}
\usepackage{url} 

\usepackage[T1]{fontenc}
\usepackage[utf8]{inputenc}

\newcommand{\blind}{1}

\begin{document}


\def\spacingset#1{\renewcommand{\baselinestretch}%
{#1}\small\normalsize} \spacingset{1}


\if1\blind
{
  \title{\bf Estimating the number of entities with vacancies using administrative and online data}
  \author{Beręsewicz Maciej\thanks{
  The authors thank Kamil Wilak, Joanna Napierała, Teresa Królikowska, Leszek Kozłowski, Matyas Meszaros, Grzegorz Grygiel and Prajamitra Bhuyan for their detailed comments on earlier drafts.
  The views expressed in this article are those of the authors and not of Statistics Poland, the Statistical Office in Poznań or the Statistical Office in Bydgoszcz.
    }\hspace{.2cm}\\
    Poznań University of Economics and Business \\ 
    Statistical Office in Poznań \\
    and \\
    Herman Cherniaiev \\
    University of Information Technology and Management in~Rzeszów \\
    and \\
    Robert Pater \\
    University of Information Technology and Management in~Rzeszów \\
    Educational Research Institute in Warsaw
    }
  \maketitle
} \fi

\if0\blind
{
  \bigskip
  \bigskip
  \bigskip
  \begin{center}
    {\LARGE\bf Estimating the number of entities with vacancies using administrative and online data}
\end{center}
  \medskip
} \fi


\bigskip
\begin{abstract}

In this article we describe a study aimed at estimating job vacancy statistics, in particular the number of entities with at least one vacancy. To achieve this goal, we propose an alternative approach to the methodology exploiting survey data, which is based solely on data from administrative registers and online sources and relies on dual system estimation (DSE). 

As these sources do not cover the whole reference population and the number of units appearing in all datasets is small, we have developed a DSE approach for negatively dependent sources based on a recent work by Chatterjee and Bhuyan (2020). To achieve the main goal we conducted a thorough data cleaning procedure in order to remove out-of-scope units, identify entities from the target population, and link them by identifiers to minimize linkage errors. We verified the effectiveness and sensitivity of the proposed estimator in simulation studies.

From a practical point of view, our results show that the current vacancy survey in Poland underestimates the number of entities with at least one vacancy by about 10-15\%. The main reasons for this discrepancy are non-sampling errors due to non-response and under-reporting, which is identified by comparing survey data with administrative data.

\end{abstract}

\noindent%
\vfill

\newpage

\spacingset{1.45} 

\section{Introduction}
\label{sec:intro}
 Statistics about job vacancies are important, yet notoriously hard to produce. From an economic point of view, job vacancies represent the portion of labor demand that has not been filled yet. After complementing the level of employment with vacancy statistics, it is possible to fully measure labor demand. Vacancy statistics have other useful functions as well. Vacancies can provide information about the skills gap experienced by companies or other structural characteristics, such as job contract types. Since vacancies represent an unmet demand, they can be treated as predictors of labor demand and can be used to construct a~leading index for the labor market. At the same time vacancies can show various mismatches that occur in the labor market and explain trends in the structural unemployment. However, vacancies are not reported for tax purposes, even though employment is. As a result, 'vacancies' are an intangible measure.

The main source of European official statistics about vacancies is the Job Vacancy Survey (JVS), conducted quarterly in all EU member states. However, job advertisements are commonly used as proxies of job vacancies. The earliest attempt to track job offers, called the Help-Wanted Index (HWI), was created in 1951 and restructured in 1987. For a long time it was based only on the classified pages of newspapers \citep{abraham1987help}. However, that dataset included only few variables. Only recently has there been a significant rise in the number of publications on how to extract more detailed information from online job offers, driven by the development of web-scraping and natural language processing methods, as well as the growing availability of open source software. Since 2005 the Conference Board publishes the Help Wanted OnLine, which is based on online job postings. Researchers who analyse job offers collect data either from one or from many websites. The advantage of the first approach is the more in-depth knowledge of a smaller amount of scrupulously classified data. The latter approach makes it possible to obtain a~lot of information with potentially many variables, but with larger mis-classification error due to automatic procedures used for text classification.

In this study we focus on job vacancies in Poland. These data come from three distinct sources. The first one is the Demand for Labor (DL) survey, a~complex random sample survey of businesses, conducted quarterly by Statistics Poland (in particular by the Statistical Office in Bydgoszcz). Its methodology is the same as that used in the pan-European JVS. The second source is the Central Job Offers Database (CBOP), which contains all job offers registered by public employment offices (PEOs), regional employment offices (REOs) and voluntary labor corps (VLCs) in Poland\footnote{In the end, we relied almost exclusively on data from PEOs, because the other sources contained foreign job offers.}. Inclusion in the database means that a~job offer was officially submitted by a~company to a PEO (or was acquired by a~PEO from a REO or a VLC), and that a~job offer was registered by completing a~form with required details. The CBOP database does not contain all job offers in Poland, but those that are included have been classified and checked by a~PEO worker. The third database contains job offers from an online job board Pracuj.pl (hereinafter Pracuj).
Websites generally provide a~richer (in terms of description) source of job postings than employment offices. However, online job offers appear on websites with various structures. They are not checked by clerks, as is the case with offers posted in PEOs, but instead are classified automatically using rule-based or machine learning algorithms. Pracuj is arguably the most recognizable job search websites in Poland \citep{megapanel}. It specializes in advertising job offers and provides detailed information about vacancies obtained from  employers. In order to post a~job offer a~company needs to pay a~fee.

In this study we use the same definition of the target population as in the DL survey but we focus on economic entities with job vacancies (i.e. legal units and their local units together; see section~\ref{sec-data} and Appendix~\ref{appen-defs} for exact definitions). In order to correctly identify units from the target population we established collaboration with the Statistical Office in Bydgoszcz (Poland), which is responsible for conducting the DL survey, to verify whether a~given entity belongs to the target population. Our target quantity is \textit{the number of units with at least one vacancy}.

Unfortunately, CBOP and Pracuj.pl do not fully cover the target population; neither are they random samples, which is why they cannot be directly used to estimate the target variable. To~overcome this, we use dual system estymation (DSE, i.e. capture-recapture methods, CR), which make it possible to estimate the target quantity after integrating the two sources. CR methods are widely used in official statistics and are recommeded as a~standard tool for assessing census quality \citep[cf.][]{un-dse}. CR methods have several assumptions, the most important being independence of data sources and absence of linkage and over-coverage errors \citep{Wolf2019, Zhang2019}. 

Our contribution can be summarized as follows. From the methodological point of view, we extend the model proposed by \citet{Chatterjee2019} and derive a point and variance estimator for negatively dependent sources. We verify performance and sensitivity analysis of the proposed method by conducting simulation studies. As regards the aspect of application, we contribute to the literature on job vacancy statistics by providing a~novel method based solely on non-survey data sources by combining administrative data with online data. We focus on the quality aspects by providing rigorous data cleaning procedures and conducting clerical review of a selected sample of job offers to verify the linkage between companies and their identifiers. We believe that the proposed method can be applied in other countries where such data sources exist.

The article has the following structure. Section~\ref{sec-back} describes international experiences regarding the production of job vacancy statistics and the use of job offers as vacancy proxies. Section~\ref{sec-data} describes the data used in the study and the procedure for preparing and cleaning online data. Section~\ref{sec-methods} provides an overview of capture-recapture methods used in the study. Section~\ref{sec-results} presents results from the study and compares them with existing results from the probability sample. Finally, section~\ref{sec-concl} summarizes the article and discusses further research steps. Additional details and results are presented in the Appendix. 


\section{Background literature}\label{sec-back}

\subsection{The definition of a job vacancy}

\textit{Eurostat}\footnote{Job vacancy statistics; \url{https://ec.europa.eu/eurostat/cache/metadata/en/jvs_esms.htm}} defines a~job vacancy as~a~\textit{paid post that is newly created, unoccupied, or about to become vacant under two conditions}:

\begin{enumerate}
  \item \textit{the employer is taking active steps and is prepared to take further steps to find a~suitable candidate for a~job from outside the enterprise concerned; and}
  \item \textit{the employer intends to fill the job position either immediately or within a~specific period}.
\end{enumerate}

This definition is similar to others, but sometimes additional constraints are included. For example \citet{holt1966concept} added a condition that a~job vacancy should include employee requirements, and, above all, the wage rate. \citet{jackman1989vacancies} reported the UK definition, which was very similar, but specified that the company should have taken a~recruiting action within four weeks before posting a~vacancy. The definition used by \textit{the Bureau of Labor Statistics}\footnote{Job Openings and Labor Turnover Survey; \url{https://www.bls.gov/opub/hom/jlt/pdf/jlt.pdf}} in the Job Openings and Labor Turnover Survey (JOLTS)  contains information about possible types of contracts: full-time, part-time, short-term, permanent, or seasonal, but the requirement is that a job should  start within 30 days. The BLS definition of a job opening excludes internal transfers of employees (promotions, demotions) and recalls of workers from layoffs. Also excluded are positions that will be filled by employees from leasing companies, temporary help agencies, outside contractors, and consultants. \textit{The Australian Bureau of Statistics} states that a vacancy must be available for immediate filling. The category excludes\footnote{Job Vacancies, Australia methodology, \url{https://www.abs.gov.au/methodologies/job-vacancies-australia-methodology/nov-2020}}:

\begin{itemize}
    \item ``jobs not available for immediate filling on the survey reference date;
    \item jobs for which no recruitment action has been taken;
    \item jobs which became vacant on the survey date and were filled on the same day;
    \item jobs of less than one day's duration;
    \item jobs only available to be filled by internal applicants within an organisation;
    \item jobs to be filled by employees returning from paid or unpaid leave or after industrial disputes;
    \item vacancies for work to be carried out by contractors; and
    \item jobs for which a~person has been appointed but has not yet commenced duty.''
\end{itemize}

Since the methodology of the Polish Labor demand survey is the same as that used in the job vacancy survey conducted by all EU member states, Eurostat’s definition of a~job vacancy also applies to this survey. Similarly to the US statistics, the Polish DL does not include B2B contracts or workers appointed for new positions within the same organisation. Unlike employment statistics, job vacancies do not include information on self-employed persons.

Job advertisements can include job positions that are not treated as employment according to labor law. In Europe, generally, the following job positions are not treated as employment contracts and are must be excluded from vacancy statistics:

\begin{itemize}
    \item independent contractors (freelance contractors);
    \item business-to-business contracts (B2B);
    \item temporary employee staffing agreements (temporary staffing agreements, temporary services agreements);
    \item apprenticeships, traineeships;
    \item voluntary workers;
    \item contracts for specific work (which can informally be called ‘work contracts’, so possibly they might be mistaken for ‘employment contracts’);
    \item contracts of mandate (in Poland);
    \item contracts for the provision of independent services;
    \item association contracts (in Greece);
    \item representation contracts (in Greece).
\end{itemize}

An employer can take various 'active steps' to find an active candidate. Out of seven active steps listed by Eurostat, there are some that are connected to job offers. Thse include:

\begin{enumerate}
  \item \textit{notifying public employment services about a job vacancy; and}
  \item \textit{advertising the vacancy in the media (for example online, in newspapers or magazines).\footnote{A `job advertisement' is a~notice containing a~`job offer' for one (most commonly) or more job positions.}}
\end{enumerate}

Other methods include using a human resources (HR) agency, approaching and recruiting a~worker directly, using personal contacts or internships, and advertising the vacancy on a~public notice board.

In this article, we use the same definition of a job vacancy as that adopted by Statistics Poland and Eurostat, for the purpose of processing and cleaning data obtained from administrative and online sources. Despite some international differences between definitions of a job vacancy, the following conditions must be met for a~workplace to be regarded as a~job vacancy: i) the workplace is unoccupied, ii) the company is looking for an employee to fill it, and iii) the company is willing to fill the position as soon as they find a suitable candidate. These conditions correspond to the definition of an unemployed person, who does not have a~job but is actively searching for one and is willing to start working as soon as they find a~suitable job. Problems occur when one attempts to to empirically measure the number of vacancies. This issue will be considered in the next section.

\subsection{Job offers as a measure of vacancies}

Online data can supplement official statistics, but it is important to remember that they are non-probability samples. \citet{beresewicz2021inferring} indicate and summarize various biases that can occur when online job offers are used as a measure of vacancies. They also point out certain advantages of using online job offers. So far, no article has been published that presents a~thorough analysis of how to address all of these biases. The solutions presented in the literature do not propose a full statistical procedure for estimating the population of job vacancies based on job offers.

\citet{carnevale2014understanding} estimate that in the US economy the share of online job offers in all vacancies in 2014 was between 60\% and 70\%, but they treat these figures as “back-of-the-envelope” estimates. \citet{acemoglu2020ai}, \citet{deming2020earnings}, \citet{forsythe2020labor}, \citet{blair2020structural}, and \citet{modestino2020upskilling} use the Burning Glass Technologies (BGT) data, which are compiled using job offers collected from many US websites. Jobs in their database accounted for 85\% of jobs from the probability sample survey (Job Openings and Labor Turnover Survey, JOLTS) in 2016. However, there is little information about the procedure of data collecting, cleaning, and classifying. Representativeness of these data was analysed using an approach similar to that adopted in previous studies by \citet{hershbein2018recessions} and \citet{deming2018skill}. The procedure involves cross-validating the data (e.g. on skills) with other measures obtained from a~probability-sample survey and then comparing the distributions of online job offers used, or their subsample, with the results of JOLTS and with the distributions of other measures of online job offers. For example, \citet{deming2020earnings} exclude vacancies where information about the employer is missing. \citet{scrivner2020job} exclude any unclassified job postings. \citet{blair2020structural} weight data by the size of the labor force and the share of employment by occupation and metropolitan statistical area. They use six-digit occupational codes in their study and include fixed effects for occupation, region, and firm. Using comparable data, \citet{shen2020measuring} conduct a similar analysis for the Australian vacancy market. 

\citet{marinescu2020opening} and \citet{marinescu2018mismatch} use a~different approach and collect data about job offers from one US website (CareerBuilder). This website provided the authors with many variables as well as information for job seekers. The main disadvantage of this approach is the low coverage of the population of vacancies. When they compared their collected online job offers with the JOLTS survey, they concluded that CareerBuilder.com represents 35\% of the total number of vacancies in the US economy.

The largest study of European job offers is conducted by Cedefop using data from all EU member states and a method proposed by Colombo et al. (2018). It involves the use of automatic web crawling and scraping algorithms as well as data provided on the basis of agreements. These data show promising results that could supplement official statistics on vacancies, but the research is still in progress \citep{beresewicz2021inferring}.

\citet{turrell2018using} and \citet{turrell2019transforming} use online job ads posted by firms and recruitment agencies on Reed.co.uk. These data cover 40\% of vacancies recorded in an official sample survey. Data are weighted in a~few job breakdowns to obtain distributions comparable to those observed in survey data (from the Vacancy Survey conducted by the UK's Office for National Statistics). Then texts of job adverts are transformed into time series data labelled by official classifications (Standard Occupational Classification codes). The authors also identify some biases in the survey-based study that do not exist in the case of online job offers. These include non-response bias, incomplete-response bias, overestimation of vacancies posted by large firms, and a~“frequency” bias (the survey produces quarterly data, while estimates based on online data can be obtained at higher frequencies). 

\citet{bhuller2019broadband} use data Public Employment Services data for Norway and a~major online Norwegian job board Finn.no. The use of online data in this study is limited because in the case of some vacancies there was no information about the economic entity that opened a job vacancy. However, the authors were able to rely on administrative data containing detailed information on posting dates, the duration of vacancies at firm and establishment level, and attrition due to non-response. As a result, they were able to link their data with other data sources about firms. They cross-validated this vacancy measure with the one produced by the Annual Survey of Establishment-level Vacancies using aggregated results. They conclude that administrative data show a similar time variation to that observed in survey data. \citet{hensvik2021job} use data from Platsbanken.se, the largest online job board in Sweden as a source of information on vacancies. They report that this data source contains 95\% of all vacancies measured by JVS in Sweden. They link this dataset with other data sources and infer about job search changes during the COVID-19 recession.

\section{Data description}\label{sec-data}

\subsection{Demand for Labor survey in Poland}


The Demand for Labor (DL) survey is carried out as a~complex probability sample consisting of 100,000 units.  The selection is made using stratified Poisson sampling, where the population is initially split into two groups: one containing entities with more than 9 employees (50,000), and the second, including companies with up to 9 employees (50,000). In 2018 the sampling frame contained 844,280 entities, including 111,000 local units and 733,000 entities of the national economy (in total 734,000 enities with one or no local unit).

Regarding the entities with more than 9 employees, the objective of the survey is to obtain information about selected sectors of the economy (by NACE sections) in each province (NUTS2 level regions). As a result, this part of the population is divided into 304 separate subpopulations. The sample of about 50,000 entities is allocated between particular subpopulations in such a way as to obtain approximately the same level of precision of results for each subpopulation. Units in each subpopulation are sorted in a~descending order according to the number of employees (according to information in the sampling frame). The largest units in each subpopulation are included in the survey without sampling. The target sample is allocated between subpopulations using the numerical optimisation method described in \citet{lednicki2003optimal, kozak2004optimal}. In the case of units with up to 9 employees, the main objective of the survey is to obtain precise results for 19 NACE sections.  Within these sections, units are stratified by province and selected using stratified, proportional sampling. Both types of entities report the \textit{number of job vacancies on the last day of the reporting quarter according to the Polish Classification of Occupations and Specialties 2014} by providing: the~number of vacancies newly created within a~given quarter, the~total number of vacancies at the end of the quarter, and the~number of vacancies at the end of the~quarter that companies reported to public employment offices. Detailed statistics about these variables are presented in section~\ref{sec-results}.

For this study we obtained anonymized unit-level data for 2018 from Statistics Poland. Table~\ref{tab-dl-survey} presents basic statistics regarding the data collection process and the estimated population size. The response rate was calculated as a~share of companies that reported to the initial sample size (we excluded inactive or out-of-scope units, the number of which ranged from 3,000 in 2018Q1 to 5000 in 2018Q4). In each quarter, 63\% of units on average responded,  with the response rate varying considerably depending on unit size. There was no decline in the overall response rate but the over-coverage error rose from 3\% to 5\%.

\begin{table}[ht!]
\centering
\caption{Basic statistics about the Demand for Labor survey for 2018}
\label{tab-dl-survey}
\begin{tabular}{llrrrr}
  \hline
    Measure & Unit size & 2018Q1 & 2018Q2 & 2018Q3 & 2018Q4 \\ 
  \hline
  Over-coverage error (\%) & Overall & 3.0 & 3.4 & 4.5 & 5.0 \\ 
  \hline
  Response rate (\%) & Small (to 9) & 36.2 & 34.6 & 34.8 & 34.3 \\
   & Medium (9-49) & 73.0 & 72.8 & 73.1 & 73.1 \\ 
   & Large (50+) & 94.1 & 94.1 & 94.2 & 94.0 \\ 
   \cline{2-6} 
   & Average & 63.3 & 62.5 & 62.9 & 62.7 \\ 
\hline   
 Estimated population  & Small (to 9) & 565,071 & 546,894 & 527,135 & 521,437 \\ 
 size ($\hat{N}$)  & Medium (9-49) & 162,738 & 164,001 & 159,786 & 160,563 \\ 
   & Large (50+) & 61,199 & 62,301 & 62,890 & 63,310 \\
   \cline{2-6}
   & Total & 789,008 & 773,196 & 749,811 & 745,310 \\
   \hline
\end{tabular}
\end{table}

The over-coverage error is the reason why the estimated population size differs from the initial figure of 884,000. Such discrepancies might have happened when a sampled economic unit ceased to exist soon before or during the survey. According to the survey, there were 789,000 and 745,000 units in the first and the last quarter of 2018 respectively. As expected, most units were small (about 70\%) and had the highest rate of non-response (about 65\%). We use this estimated population size as a~reference for our study.

\subsection{Administrative data from public employment offices}

Public employment services in Poland are made up of public employment offices (PEOs), which operate at NUTS4 level (\textit{ Nomenclature des unités territoriales statistiques}, \textit{Nomenclature of Territorial Units for Statistics}) and are responsible for registering and managing unemployment. This means that there are two sources of data about employment in Poland: (i) the labor force survey (LFS) and (ii) registered data collected by PEOs. The register maintained by PEOs contains detailed characteristics of unemployed persons and job offers. The number of registered unemployed may be higher than the LFS estimate because registration is connected with free health insurance and unemployment benefits. 
The number of job offers is known to be lower than that of vacancies, because not every company submits their job openings to PEOs. Although this information should be reported according to the 2004 act on promotion of employment and labor market institutions, many companies fail to do so in the absence of legal consequences. \citet{gradzewicz2016badanie} show that the most important source of job vacancies was the media (approx. 50\% of all vacancies). The second  most frequently used way of disseminating information about vacancies was recommendation (approx. 30\%). Only about 10-20\% of all job offers are submitted to PEOs, and over 60\% of all employers do not send their offers there at all. Data for 2018-2019 from the Balance of Human Capital Study reveals that 64\% of companies adertised their openings through PEOs in 2018, compared to 51\% in 2019, 47\% used job boards in 2018 and 42\% in 2019, 41\% relied on recommendations from current employees in 2018 and also in 2019, and 33\% of companies used internal job postings in 2018, compared to 35\% in 2019. According to the 2018 DL survey, about 25\% of entities reported vacancies to PEOs, which constitutes about 20\% of all vacancies.

The structure of job offers from PEOs is different than that observed in the DL survey. For one thing, there may be an over-representation of jobs from companies that have an incentive to advertise their vacancies through public employment offices, for example in the case of refunded internships or publicly-subsidised workplaces for the disabled. Public entities, in particular, are more willing to publish job offers in PEOs because they are often obliged to do so by their own internal regulations. Finally, low-paying jobs are more often sent to PEOs because people with lower qualifications often rely on public institutions to help them find a job. Better-paying jobs are more often advertised on job boards, in media or through private HR agencies, which charges fees for their services.

However, registered data also have valuable properties. They provide information about stocks and flows of job offers. The PEO register contains structured fields with detailed workplace description, including occupation, type of contract etc. Each submitted vacancy can be classified by a~qualified PEO worker or by an employer. In the first case, one can be sure that industry and occupation is properly coded. In the second case, a~PEO worker checks whether all information has been provided and whether it looks plausible (as required by the law, see a detailed description in Appendix~\ref{peo-details}). If the employer has not included the NACE section or the ISCO occupational code, PEO staff are obligated to supplement this information based on the company tax identification number and job description\footnote{Based on interviews with PEO staff we know that the majority of employers do not provide occupational codes}. If submitted information looks suspicious, e.g. if it contains many workplaces or the format of working hours seems to be wrong, the company is contacted directly with a request for clarification. This generally guarantees a~higher quality of data compared to information obtained from commercial websites and automatic classification algorithms. 

Job offers are also verified by PEO staff in terms of their expiration date, which reduces the over-coverage error. If the employer has indicated an expiration date on their offer, it is manually removed on that day. An employer who finds a suitable employee should immediately report this fact to the PEO so that the advert can be removed, together with the reason why the offer is no longer valid, which is also registered. While this requirement is not strictly obeyed, it is verified to some extent. On the basis of information provided by employers, PEO determine the frequency of contacts with particular employers. These contacts should not be less frequent than once a~month. Usually, they happen once every two weeks. In the case of foreign job offers, the required frequency is every two weeks. After this period, a~PEO employee will contact the employer using the method and contact indicated earlier and asks if the workplace still has not been filled. Further description of the data is presented in Appendix~\ref{peo-details}.

\subsection{Commercial employment web services} 

Since CBOP is a~governmental service, we decided to use a~commercial employment job board as an additional source of data about job offers. There are many websites which advertise job offers, and their number cannot be accurately determined. Moreover, such websites differ from one another in terms of various features (e.g. paid vs. free websites, occupation-specialized etc.). These websites can be divided into the following groups:

\begin{enumerate}
\item country-wide online recruitment services,
\item industry-specific websites (limited to e.g. IT or financial occupations),
\item local job search websites (e.g. limited to jobs offered within a paricular local labor market defined by various (2-5) NUTS levels),
\item employers' websites,
\item Internet forums with job offers (e.g. Facebook groups),
\item job aggregators (e.g. Jooble).
\end{enumerate}

A systematic collection of job offers is a~non-trivial task. Unlike  CBOP, online job boards usually do not have a~public API, so it is necessary to use web scraping algorithms. Another condition is that such websites contain an archive with job offers or else historical job offers cannot be accessed after they expire. 

Based on Google Trends and results of \citet{megapanel} we decided to use Pracuj.pl (Pracuj; \url{https://www.pracuj.pl/}), because it has the biggest number of users. This job board also publishes all archived advertisements (\url{http://archiwum.pracuj.pl/}). We also examined another data source -- OLX -- which is among the top 5 most frequently used online job boards in Poland according to \citet{megapanel}. However, it did not meet data quality requirements. More details about OLX are included in Appendix~\ref{online-selection-details} and~\ref{olx-service}. 
 
The biggest difference between CBOP and Pracuj is that the latter, being a~commercial website, charges a fee for posting a~job offer. Like other nation-wide job boards, Pracuj requires registration, but unlike some of them, it provides open access to employers' profiles. After opening a profile, users can track current and expired job offers.

Because collected job advertisements did not contain company tax identifiers (REGON or NIP numbers) we conducted a~multi-step data cleaning procedure involving text mining techniques, matching entity names between CBOP and Pracuj as well as automated Google search queries to identify units that belong to our target population. A detailed description of all steps is given in Appendix~\ref{pracuj-preparation}. 


\subsection{Data comparison}

Table~\ref{tab-cbop-quality} contains data quality indicators for CBOP and Pracuj databases. Both sources had a~similar number of records at the beginning\footnote{Note that each record represents a row in~a data table, which contains one advertisement.}. In the first step we selected only those ads that were available at the end of each quarter. To do that, we used variables with information about the date of initial placement, expiration date and archiving date. 

\begin{table}[ht!]
    \centering
    \caption{Data quality indicators of CBOP and Pracuj.pl for 2018}
    \label{tab-cbop-quality}
    \begin{tabular}{lrr}
    \hline
    Indicator & CBOP & Pracuj \\
    \hline
    Raw data & 516,260 & 529,447 \\
    Records at the end of quarter(s) & 121,804 & 163,790 \\
    \hline
    Records with erroneous, missing or hidden employer ids &  41 & 4,195 \\
    Records with errors in date fields & 6,781 & 6 \\         
    Records with out of scope units (not job offers, archived) & 14,327  & 10,057 \\     
    Records with out of scope units (employers) & 17,166 & 13,899 \\
    \hline
    Final number of records  & 83,695 & 135,851 \\
    \hline
    \end{tabular}   
\end{table}

In the next step, we focused on removing records with missing or erroneous data in employer fields. For instance, in CBOP there were 41 records without either REGON or NIP id, while in the case of Pracuj, the most frequent problem was missing company name (e.g. hidden recruitment) or the impossibility to assign an identifier based on the company name or Google search (see Appendix~\ref{pracuj-preparation} for details).  

Next, we removed records containing missing or erroneous values in key variables, such as the number of vacancies (CBOP), dates (Pracuj) or location (region).  The main difference between the two datasets were erroneous values in employer fields in Pracuj, and the missing number of reported vacancies in CBOP.

Finally, we removed records that did not belong to the populations of job vacancies and entities. Note that we did not deduplicate data for both data sources as we were not interested in estimating the total number of vacancies. A comparable number of records was removed from both datasets owing to over-coverage. 10,000-15,000 records turned out not to be job offers (e.g. internships, non-job contracts, such as B2B contracts) and  10,000-15,000 records referred to employers that did not occur in the reference population defined by the DL survey. After completing the cleaning procedure our two datasets contained about 84,000 (CBOP) and 136,000 records (Pracuj). The CBOP database contained only 64 local units and Pracuj -- only 47, which is significantly fewer than the number found in the sampling frame for the DL survey. That is why we decided to limit our analysis to entities of the national economy and not treat their local units separately.

Table~\ref{tab-statistics-sources} contains a comparison between the three sources used in the study. Note that the DL survey contains a question about whether a given unit has reported a vacancy to a PEO. With this information we were able to assess the non-response or under-reporting error based on administrative data about all job vacancies registered by PEOs.

\begin{table}[ht!]
    \centering
     \caption{Number of companies and vacancies according to the DL survey, CBOP and Pracuj}
    \label{tab-statistics-sources}
    \begin{tabular}{ll|rrrr}
    \hline
     Source &  Variable  & 2018Q1 &  2018Q2 & 2018Q3 & 2018Q4  \\
     \hline
     DL survey & \multicolumn{5}{c}{All} \\
            \cline{2-6}
            & Vacancies & 152,414  &  164,745  & 157,155  &   139,193  \\
            & Companies &  54,655  &   56,076 &  50,805 &   44,108    \\
            & of which: Entities &  47,509  &   46,562 &  43,575 &   38,490    \\
            & of which: Local units &  7,146   &   9,514 &  7,230 	 &   5,618  \\
            \cline{2-6}
            & \multicolumn{5}{c}{Reported to PEOs} \\
            \cline{2-6}
            
            & Vacancies  & 35,181	  &  34,132  &  30,283 &    25,337	   \\
            & Companies  &  14,577  & 15,500   & 11,719  &   9,989    \\
            & of which: Entities  &  13,072  &   11,403 &  10,232 &   9,181    \\
            & of which: Local units &  1,505  &   4,097  &  1,487 	 &   808    \\
            \hline
     CBOP & Vacancies & 55,111  &  64,917  & 59,213  &  31,981 \\
          & Entities & 14,965 &  16,420  & 15,902  & 9,475   \\
            \hline          
     Pracuj & Vacancies & 35,683 & 36,046 & 36,564 & 27,558 \\
            & Entities & 8,057 & 7,865 & 7,287 & 6,376 \\
     \hline
    \end{tabular}
\end{table}

Compared with the DL survey, CBOP provides a~biased estimate of vacancies and of the number of entities having a~vacancy for each quarter of 2018. This may be due to under-reporting (e.g. CBOP contains nearly 65,000 vacancies in 2018Q2 while the corresponding DL estimate is over 34,000) and non-response. At the moment of writing this article, we did not have access to de-anonymized unit-level data to investigate these differences. This discrepancy is another motivation for using alternative methods and sources to correctly estimate the total number of units with job vacancies.

\section{Methods}\label{sec-methods}

\subsection{Estimation of the population size using the capture-recapture approach}

Determining the number of companies with job vacancies is a~methodological challenge for official statistics. On the one hand, administrative registers, like the one created by public employment offices, or online job board services cover only a~fraction of the whole population. On the other hand, growing non-response in sample surveys can result in underestimation if auxiliary variables do not correctly account for this process. It is, therefore, necessary to develop alternative ways of deriving this estimate that rely on existing sources and reduce respondent burden.

A~number of appropriate statistical methods for estimating population sizes based on capture-recapture techniques have been proposed in the literature \citep[for a~recent review see][]{bohning2017}. The most popular ones include dual system estimation (DSE; two sources) or multiple system estimation (MSE; three or more sources), both of which are used for estimating hard-to-reach populations or register-based census statistics. These methods require access to unit-level data and are based on certain assumptions (the same probability of captures, no over-coverage, independent sources or perfect linkage), which might be difficult to meet in practice \citep[cf.][]{zaslavsky1993triple, wolter1986some, Zhang2019}. 

To overcome these issues, a~number of appropriate methods have been proposed in the literature. Stratification is used to account for between-group heterogeneity, under the assumption of within-group homogeneity \citep{cormack1989log, van2012people}. \citet{Zhang2015} considered the case of two data sources, where the first suffers from over- and under-coverage and the second (i.e. independent survey) suffers only from under-coverage. Further, \citet{zhang2017trimmed} proposed trimmed dual system estimation (TDSE), where a~certain number of records in one or both sources are removed based on a criterion.  \citet{ding1994dual, di2015coverage, di2018population, zult2019general} proposed methods to correct for linkage errors, which require clerical review to calculate the rate of false negative and false positive linkages. Finally, \citet{Chatterjee2018,Chatterjee2019a,Chatterjee2021} proposed estimators for dependent dual system estimators (DSE) and \citet{Gerritse2015, Chatterjee2020} discussed scenarios for detecting dependence in DSE. 

In this article we extend the method proposed by \citet{Chatterjee2019a} assuming a~negative correlation between the sources and truncation of data based on the number of days prior to the end of the quarter.

\subsection{Dual system estimation based on independent sources}

The starting point for DSE is an contingency table containing information from two sources, as shown in Table~\ref{tab1}.  In the case of two data sources A~and~B, there may be a~situation where, after the units are combined, there are only units in~source A and~not in~source B, denoted by $n_{10}$, or only in~source B and not in~source A, denoted by $n_{01}$, or units occur simultaneously in~sources A~and~B, denoted by $n_{11}$. The objective is therefore to estimate the number of $n_{00}$, i.e. the number of units not included in either of the sources. The final estimated population size is obtained by adding all the values from Table~\ref{tab1} having first estimated the size of $n_{00}$.

\begin{table}[ht]
\centering
\caption{The case of two sources -- $2\times 2$ contingency table}\label{tab1}
\begin{tabular}{ll|ll|l}
\hline
 &  & \multicolumn{2}{c|}{List 2} &   \\ 
  &  & Yes (1) & No (0) & $\sum$\\ 
 \hline
List 1 & Yes (1) & $n_{11}$ ($p_{11}$) & $n_{10}$ ($p_{10}$) & $n_{1.}$ ($p_{1.}$) \\
 & No (0) & $n_{01}$ ($p_{01}$) & $n_{00}$ ($p_{00}$) & $n_{0.}$ ($p_{0.}$) \\ 
\hline
$\sum$ & & $n_{.1}$ ($p_{.1}$) & $n_{.0}$ ($p_{.0}$) & $n$ ($p_{..}$) \\ 
\hline
\end{tabular}
\label{tab-cont}
\end{table}

In order to estimate the total population size we can use the Lincoln-Petersen estimator, given by the following equation:

\begin{equation}
    \hat{N}_{\text{na\"ive}} = \frac{n_{1.}n_{.1}}{n_{11}}.
    \label{cr-petersen}
\end{equation}

In this article we will refer to this estimator as \textit{the na\"ive estimator}, since it is based on assumptions that may not hold in real data applications.

\subsection{Proposed dual system estimation for dependent sources}

In the setting with two dependent sources two scenarios can be considered: (1) a~positive dependence, where units are more likely to be observed in the second source/time; and (2) a~negative dependence, where units are less likely to be observed in the second source/time. In the literature, these scenarios are often referred to as \textit{behavioral response effects}. The first scenario is observed in post-enumeration surveys used for assessing census under-count (see \citet{bell1993using} for USA and \citet{chatterjee2016improved} for India), while the second can be observed in situations where both sources are exclusive (e.g. children injury data collected by hospitals and police stations) or re-identification is associated with social stigma (e.g. the population of drug users, patients infected with HIV).  In such situations, the Lincoln-Petersen estimator, given by~\eqref{cr-petersen}, is biased and \citet{Chatterjee2021} showed that the approximate bias is given by

\begin{equation}
    \text{Bias}(\hat{N}) \approx N(1-p_{1.})\frac{1-\phi}{\phi} + \frac{1}{\phi} \frac{(1-p_{1.})(1-\phi p)}{p_{1.}\phi p},
\end{equation}

\noindent where $\phi > 0$ denotes a \textit{behavioural response effect}, $p = p_{01}/p_{0.} = p_{01}/(1-p_{1.})$ is Pr(an individual is captured in List 2 $\mid$ not captured in List 1), $p_{1.}$ is probability of being captured in List 1 and $N$ is the population size. 

To verify whether two sources are dependent, \citet{Chatterjee2020} suggest calculating $c = p_{11}/p_{1.}$, which is Pr(an individual is captured in List 2 $\mid$ captured in List 1) and comparing it with $p$ to verify whether there is $\phi$ that $c=\phi p$. The negative dependence would be represented by smaller values of $c$ (closer to 0), while the positive dependence -- by larger values of $c$ (closer to 1). In the article, we calculate $c$ to indicate the direction of the relationship. Below we focus on the method proposed by \citet{Chatterjee2019a} to estimate the population size under dependence. 

Let us consider population $U$ of size $N$ and let $p_{j1.}, p_{j.1}$ denote the capture probabilities of the $j$-th individual in the first ($Y$) and in the second ($X$) list. Let $\alpha$ be a~proportion of individuals for whom there is a behavioural dependence between List 1 and List 2. Let $Y$ denote inclusion in List 1, and $Z$ inclusion in List 2. To capture the dependency structure we define a~pair $(X^*_{1j},X^*_{2j})$, which represents the latent capture status of the $j$-th individual ($j=1,...,N$) during the first and second attempt. The latent capture status $X_{lj}$ takes values $\{0,1\}$, which denote the absence or presence of the $j$-th individual in the $l$-th list $(l=1,2)$.

In this approach we assume that $\alpha$ is the proportion of individuals for whom there is a behavioural dependence i.e. the value $X^*_{2j}$ is the same as of $X^*_{1j}$ (i.e. $X^*_{2j}=X^*_{1j}$). Now, note that $(Y_j,Z_j)$ is the manifestation of the latent structures ($X^*_{1j}, X^*_{2j}$) for the $j$-th individual. Therefore, the positive dependence  between sources can be formulated as follows

\begin{equation}
\left(Y_{j}, Z_{j}\right)=
\begin{cases}
\left(X_{1 j}^{*}, X_{2 j}^{*}\right) & \text { with prob. } 1-\alpha, \\ \left(X_{1 j}^{*}, X_{1 j}^{*}\right) & \text { with prob. } \alpha,
\end{cases}
\label{model-pos}
\end{equation}

\noindent where, $X_{1 j}^{*}$ and $X_{2j}^{*}$ are independently and identically distributed Bernoulli random variables with parameters $p_1$ and $p_2$. For the negative dependence between sources the relationship can be represented as follows

\begin{equation}
\left(Y_{j}, Z_{j}\right)=
\begin{cases}
\left(X_{1 j}^{*}, X_{2 j}^{*}\right) & \text { with prob. } 1-\alpha, \\ \left(X_{1 j}^{*}, 1-X_{1 j}^{*}\right) & \text { with prob. } \alpha.
\end{cases}
\label{model-neg}
\end{equation}

\noindent Now, we can denote $\text{Pr}(Y=y,Z=z)$ under model~\eqref{model-neg} for the contingency table in Table~\ref{tab-cont}

\begin{equation}
    \begin{aligned}
    p_{11} & = (1-\alpha)p_1 p_2, \\ 
    p_{10} & = \alpha p_1 + (1-\alpha) p_1 (1-p_2) = p_1(\alpha + (1-\alpha)(1-p_2)), \\
    p_{01} & =  \alpha (1-p_1) + (1-\alpha)(1-p_1)p_2 =(1-p_1)(\alpha - (1-\alpha)p_2), \\
    p_{00} & = (1-\alpha)(1-p_1)(1-p_2),
    \end{aligned}
    \label{eq-model1-probs}
\end{equation}

\noindent where these probabilities are calculated using the law of total probability, that is 

\begin{equation}
    \begin{aligned}
    \text{Pr}(Y = i , Z = i) & = \text{Pr}(Y = i, Z = i| X_1 = i, X_2 = i) \text{Pr}(X_1 = i, X_2 = i) \\
    & + \text{Pr}(Y = i, Z = i| X_1 = i) \text{Pr}(X_1 = i).
    \end{aligned}
\end{equation}

\noindent For example, $p_{11}$ is calculated as follows

\begin{equation}
    \text{Pr}(Y = 1, Z = 1) =  p_1 \times p_2 \times (1-\alpha) + 0 \times \alpha = p_1 p_2 (1-\alpha),
\end{equation}

\noindent because in the case of negative dependence $\text{Pr}(X_{1 j}^{*}, 1-X_{1 j}^{*})$ is 0, as $\alpha$ denotes the probability of shared units that are in one but not in other list. The corresponding marginal probabilities are given by

\begin{equation}
    p_Y = p_{1.} = p_1, \quad p_Z = p_{.1} = \alpha p_1 + (1-\alpha) p_2,
\end{equation}

\noindent with $Cov(Y,Z)=\alpha p_1(1-p_1)$.

The parameters of the model~\eqref{model-neg} can be practically interpreted, i.e $\alpha$ represents the share of behaviourally dependent individuals in the population and $p_l$ is the capture probability of a~causally independent individual in the $l$-th list.

However, the parameters of this model are not identifiable as their number exceeds the observed counts. To overcome this problem, we assume that population $U$ can be stratified into two, mutually exclusive and exhaustive sub-populations, denoted by $U_A$ and $U_B$.

In addition to the standard assumptions of capture-recapture, i.e. 1) the population is closed and 2) the probability of capture in each of the two attempts is the same, we specify the following assumptions that underlie model 2 proposed by \citet{Chatterjee2019a}:

\begin{enumerate}
    \item initial (List 1) probabilities of capturing individuals belonging to both sub-populations are the same (i.e. $p_{1.A}=p_{1.B}=p_1$), which implies that $N_A p_{1A} = N_A p_1 =x_{1.A}$, $N_A p_{1B}=N_A p_{1}=x_{1.B}$ thus $N_A = (x_{1.A}/x_{1.B})N_B$,
    \item probability $p_2$ differs between populations i.e. there are two parameters $p_{2A}$ and $p_{2A}$. We can establish a relationship between $p_{2A}$ and $p_{2B}$ based on the method of moments, starting with the following set of equations based on~\eqref{model-neg}

\begin{equation}
    \begin{aligned}
    N_A(1-\alpha) p_1 p_{2A} & = x_{11A} \\
    N_B(1-\alpha) p_1 p_{2B} & = x_{11B}, 
    \end{aligned}
\end{equation}

\noindent and keeping in mind that $N_A = (x_{1.A}/x_{1.B})N_B$ we get

    \begin{equation}
    p_{2A} = \frac{x_{11A}}{x_{11B}} \frac{x_{1.B}}{x_{1.A}}p_{2B}. 
    \label{mme-p2a}
\end{equation}
\end{enumerate}

Under these assumptions, we need to estimate 6 parameters: $\mathbf{\theta} = (N_A, N_B, \alpha, p_1, p_{2a}, p_{2b})$ but this number can be reduced as $N_{A} =  (x_{1.A}/x_{1.B})N_B$ and because of the relation between $p_{2a}$ and $p_{2b}$.

In the next section we discuss how to estimate these parameters using the maximum likelihood method.

\subsection{Derivation of the proposed estimator}

\subsubsection{Maximum likelihood method}

For a~positively dependent estimator under the latent structure given by~\eqref{model-pos}, \citet{Chatterjee2019a} derived a~closed form of the method of moments estimator (MME) and then used it as the starting point for the maximum likelihood estimator (MLE), which is minimized by the optimization procedure using the Broyden–Fletcher–Goldfarb–Shanno (BFGS) algorithm implemented in the \texttt{optim} function in the R language \citep{r-cran}.

For the model given by~\eqref{model-neg} we decided to use a~constrained MLE that maximizes the following likelihood function~\eqref{model2-ll}

\begin{equation}
\begin{aligned} L\left(\mathbf{\theta} \mid \mathbf{x}_{A}, \mathbf{x}_{B}\right) \propto & \frac{N_{A} ! N_{B} !}{\left(N_{A}-x_{0 A}\right) !\left(N_{B}-x_{0 B}\right) !} \\
& \times \left[ (1-\alpha)p_1 p_{2A} \right]^{x_{11A}} \times \left[ (1-\alpha)p_1 p_{2B} \right]^{x_{11B}} \\
& \times \left[\alpha p_1 + (1-\alpha)p_1 (1-p_{2A}) \right]^{x_{10A}} \times \left[\alpha p_1 + (1-\alpha)p_1 (1-p_{2B}) \right]^{x_{10B}} \\
& \times \left[\alpha (1-p_1) + (1-\alpha)(1-p_1)p_{2A} \right]^{x_{01A}} \times \left[\alpha (1-p_1) + (1-\alpha)(1-p_1)p_{2B} \right]^{x_{01B}} \\
& \times\left[(1-\alpha)(1-p_1)(1-p_{2A})\right]^{\left(N_{A}-x_{0 A}\right)} \\ & \times\left[(1-\alpha)(1-p_1)(1-p_{2B})\right]^{\left(N_{B}-x_{0 B}\right)},
\end{aligned}
\label{model2-ll}
\end{equation}

\noindent under the following constraints

\begin{equation}
    \begin{aligned}
        N_A & \in [x_{0A}, \hat{N}_{A, \text{na\"ive}}], \\
        N_B & \in [x_{0B}, \hat{N}_{B, \text{na\"ive}}], \\
        p_{1}, \alpha, p_{2A}, p_{2B} & \in [0,1],
    \end{aligned}
\end{equation}

\noindent where $\mathbf{\theta} = (N_A, N_B, \alpha, p_1, p_{2a}, p_{2b})$. In the constrained log-likelihood we defined the lower and upper bounds of population size for $N_A$ and $N_B$. The lower bounds are equal to the observed counts, denoted by $x_{0A}$ and $x_{0B}$ respecitvely. 




In the actual computation we minimize the negative log-likelihood function, which requires the calculation of $\log(N_A!)$ etc. To overcome this, we used the approximation $\log(N_A!) \approx N_A\log(N_A) - N_A$. For more details, see Appendix~\ref{app-model2-mle}.

Because~\eqref{model2-ll} is sensitive to the starting points, we decided to use the non-linear \texttt{Ipopt} optimizer \citep{wachter2006implementation} implemented in the \texttt{JuMP.jl} module  \citep{DunningHuchetteLubin2017} available in the \texttt{Julia} language \citep{Julia-2017}. 


\subsubsection{Variance estimation}\label{sec-variance-est} 

In the article, we calculated standard errors using two techniques. First, we derived a~Hessian for log-likelihood \eqref{model2-ll} and evaluated it at $\hat{\theta}$ to obtain standard errors of the parameters. We also compared the Hessian with the forward mode automatic differentiation implemented in the \texttt{ForwardDiff.jl} module \citep{RevelsLubinPapamarkou2016} to evaluate it at $\hat{\theta}$. To calculate the variance of the total size $N=N_A+N_B$, we added variance for strata $A$ and strata $B$, as these sub-populations are independent. 

Then, we used non-parametric bootstrap (also called \textit{imputed bootstrap}, see \citet[p. 8]{bohning2019identity}) consisting of the following steps:

\begin{enumerate}
    \item draw, independently for each stratum $s=A,B$, a sample of $(x_{11s}^*,x_{10s}^*,x_{01s}^*,x_{00s}^*)$ of total size $\hat{N}_s$ with probabilities calculated as $(p_{11s}=x_{11s}/\hat{N}_s,p_{10s}=x_{10s}/\hat{N}_s,p_{01s}=x_{01s}/\hat{N}_s,p_{00s}=\hat{x}_{00s}/\hat{N}_s)$,
    \item derive $\hat{N}_s$ from the model given by~\eqref{model2-ll},
    \item repeat 1) and 2) $B$ times to obtain $N_s^{(1)},...,N_s^{(b)}$ estimates,
    \item calculate the bootstrap standard error as 
    $$
    SE^{*} = \sqrt{\frac{1}{B}\sum_{b=1}^B(N_s^{(b)} - \bar{N}_s^{*})^2},
    $$
    where $\bar{N}_s^{*} = \frac{1}{B}\sum_{b=1}^B N_s^{(b)}$.
\end{enumerate}

For the total $N$ we derived variance in the same way but instead of $N_s^{(b)}$, we used $N^{(b)}=N_A^{(b)}+N_B^{(b)}$. 

We also considered a 95\% confidence interval suggested by \citet[p. 11]{Chatterjee2019a}

\begin{equation}
\left[
x_{0s}+\left(\hat{N}_{s}-x_{0 s}\right) / C, 
x_{0s}+\left(\hat{N}_{s}-x_{0 s}\right) C,
\right]
\label{eq-ci-proposed}
\end{equation}

\noindent where $s$ denotes a stratum, $C=\exp\left\{ 1.96 \sqrt{\log( 1 + \sigma^2_{N_s} / (\hat{N_s} - x_{0s})^2)} \right\}$, and $\sigma^2_{N_s}$ is the estimated variance of $\hat{N}_s$ calculated using either the Hessian or the bootstrap approach. 

We conducted simulation studies to compare the nai\"ve approach given by~\eqref{cr-petersen} with the proposed estimator obtained from the likelihood function given by~\eqref{model2-ll}. In the first study, we focused on the properties of estimators by evaluating the bias and coverage of the confidence intervals under the model assumptions. In the second one, we verified the bias when the main identification assumption, i.e. $p_{1A}=p_{1B}=p_1$, is not met. Thses simulations show that the estimator is unbiased if $p_{1A}$ is close to $p_{1B}$ but as the difference increases, so does the bias. Given the constraints, the bias of the proposed method is lower than that of the nai\"ve approach. For details, please consult Appendix~\ref{sec-sim-study}.

\section{Results}\label{sec-results}

\subsection{Linkage and over-coverage}

Table~\ref{tab-statistics-linkage} contains descriptive statistics about the DL sampling frame (denoted as 'Frame'), the DL survey results averaged over the whole period (denoted as 'Survey'), and unique entities identified in CBOP and Pracuj for 2018. We present information about the number of entities (in thousands) and proportions by sector of ownership, size and selected NACE sections.

\begin{table}
\centering
\caption{Distribution of auxiliary variables in the DL sampling frame and units identified in the non-probability sources for 2018 (in thousands)} 
\label{tab-statistics-linkage}
\begin{tabular}{ll|rr|rr|rr|rr|}
  \hline
  & & \multicolumn{2}{c}{Frame} & \multicolumn{2}{c}{Survey} & \multicolumn{2}{c}{CBOP} & \multicolumn{2}{c|}{Pracuj}\\
  \cline{3-10}
Variable & Level & N  & \% & N & \% & N & \% & N & \% \\ 
  \hline
  Sector & Public  & 67.1 & 9.0 & 4.8 & 9.4 & 4.7 & 11.8 & 0.6 & 3.7\\ 
   & Private & 678.2 & 91.0 & 46.6 & 90.6 & 35.2 & 88.2 & 15.8 &  96.3\\ 
   \hline
  Size & Large (over 49) & 63.3 & 8.5 & 8.2 & 15.9 & 8.4 & 21.1 & 7.0 & 42.7\\ 
   & Medium (10--49) & 160.5 & 21.5 & 13.3 & 25.8 & 13.4 & 33.6 & 5.4 & 32.9\\ 
   & Small (up to 9) & 521.5 & 70.0 & 30.0 & 58.3 & 18.1	 & 45.4 & 4.0& 24.4\\  
   \hline
  NACE & C & 94.3 & 12.7 & 9.4 & 18.3 & 8.9  & 22.4 & 4.0 & 24.5\\ 
  (selected) & F & 81.6 & 5.5 & 9.12 & 17.7 & 13.8 & 12.2 & 1.0 &	6.1 \\  
            & G & 232.7 & 31.2 & 11.3 & 22.0 & 8.8 & 22.1 & 4.1 & 25.2 \\ 
           & H & 46.4 & 6.2 & 4.7 & 9.0 & 2.2  & 5.5 & 0.8 & 4.9\\ 
           & I & 36.7 & 4.9 & 2.8 & 5.4& 2.5  & 6.3 & 0.3 & 1.8 \\ 
           & J & 15.0 & 2.0 & 1.2 & 2.4 & 0.4  & 1.0 & 1.4 &	8.6 \\ 
           & K & 14.9 & 2.0 & 0.5 & 1.0 & 0.4  & 1.0 & 0.6 &	3.7\\ 
           & L & 14.8 & 2.0 & 0.6 & 1.1 & 0.5  & 1.3 & 0.4& 2.5\\ 
           & M & 53.7 & 7.2 & 2.6 & 5.0 & 1.5  & 3.8 & 2.0& 12.3 \\  
           & N & 19.2 & 2.6 & 1.2 & 2.3 & 1.4  & 3.5 & 0.8 & 4.9\\ 
           & O & 6.9 & 0.9 & 1.4 & 2.7 & 0.8  & 2.0 & 0.1 &	0.6\\  
           & P & 48.2 & 6.5 & 1.6 & 3.2 & 2.4  & 6.0 & 0.2& 1.2\\ 
           & Q & 33.6 & 4.5 & 2.0 & 3.9 & 1.9  & 4.8 & 0.2 & 1.2\\ 
           & R & 10.1 & 1.4 & 0.5  & 1.0 & 0.4  & 1.0 & 0.1& 0.6\\  
           & S & 20.5 & 2.8 & 1.6 & 3.2 & 1.0  & 2.5 & 0.1 & 0.6\\ 
   \hline
\end{tabular}
\begin{flushleft}
\scriptsize
Note: C -- Manufacturing; F -- Construction; G -- Trade; repair of motor vehicles; H -- Transportation and storage; I -- Accommodation and catering; J -- Information and communication;  K -- Financial and insurance activities; L -- Real estate activities; M -- Professional, scientific and technical activities; N -- Administrative and support service activities; O -- Public administration and defence; compulsory social security; P -- Education; Q -- Human health and social work activities; R -- Arts, entertainment and recreation; S -- Other service activities.
\end{flushleft}
\end{table}

As expected, CBOP covers more public sector entities (almost 12\%) in comparison with commercial portals, where this share is about 4\%. Both sources differ with respect to the survey, but CBOP seems to be closer. 

As far as entity size is concerned, the majority of companies in the DL population and the survey are small (70\% and 58\% respectively), while the corresponding proportions in CBOP and Pracuj are considerably smaller: 45\% and 24\% respectively. This means that the two sources underrepresent small companies, which may have limited budgets for online activities or be less willing to search for employees via administrative sources or online services.  The differences between CBOP and Pracuj suggest that both sources cover different sub-populations, which means that the information they provide is likely to be complementary. On the othjer hand, both sources contain a similar share of medium-sized companies and, compared with the survey results, overrepresent large entities.

Finally, we compared these sources across selected NACE sections. The main difference between them exists in the case of Construction (section F) and Professional, scientific, and technical activities (section M). This discrepancy is mainly due to the commercial character of Pracuj, which targets highly-skilled professionals. The distribution of companies advertising in CBOP is consistent with survey results, with small differences that may not be related to sampling and non-response error.  

Based on this analysis, it seems that the differences between the three sources are mainly associated with company size. In our estimation, we will use company size as a~post-stratification variable for the negative dependence model.

To verify over-coverage due to out-dated advertisements, we assessed the number of days between the publication date and the end of a given quarter. This result is reported in Table~\ref{tab-ads-days}. We counted the number of ads in CBOP and Pracuj according to the number of days where \textit{zero} means that the publication date is exactly the same as the end of the quarter, \textit{up~to 10} means that the ad was placed 10 days before the end of the quarter, and so on. 

We found that most ads posted on Pracuj had been placed 30 days before the quarter ends in 2018. This is because of a system used by Pracuj, where companies pay for their ads to be posted for 30 days, upon which time they are automatically archived. The situation looks different in the case of CBOP, which contains ads published over 30 days or even a year earlier. 

\begin{table}
\centering
\caption{The number of ads posted on CBOP and Pracuj depending on the number of days before the end of the quarter in 2018}
\label{tab-ads-days}
\begin{tabular}{lrr}
\hline
Days & CBOP & Pracuj\\
\hline
zero & 341 & 690\\
up to 10 & 24,553 & 33,507\\
(10,20] & 23,102 & 54,081\\
(20,30] & 14,873 & 47,455\\
(30,60] & 14,434 & 118\\
60 to a~year & 11,041 & --\\
over a~year & 61 & --\\
\hline
\end{tabular}
\end{table}

In order to harmonize the periods and minimize the over-coverage error, we decided to disregard ads posted over 30 days earlier and then we counted the number of unique entities. 

In the next step, we verified if false-negative and false-positive linkages exist by conducting a clerical review of linked units. We verified identifiers with different entity names and unique names with multiple ids but we did not find any units that had been omitted or wrongly linked. To clarify, it is possible that the sampling frame contains entities with the same name but with different ids if they operate in different regions. In addition, within each quarter we selected a sample of 50 entities from among the matches and of 150 units from the online source. Having reviewed this sample, we did not find any false-negative or false-positive linkages -- all identifiers had been assigned correctly. 


The resulting number of unique entities and estimated values of $\hat{c} = p_{11}/p_{1.}$ are presented in Table~\ref{tab-cr-data-for-model}. First, the estimated value of $\hat{c}$ is very small for both categories of companies, in particular for small \& medium-sized ones. This clearly indicates a negative dependence, which means that companies choosing to look for employees through PEOs are not willing to use Pracuj, which may be due to the fact that such services cost money and, also, their target group is different. Estimation based on the na\"ive Lincoln-Petersen estimator~\eqref{cr-petersen} will be biased because of the small number of units observed in both sources.

\begin{table}
\centering
\caption{The number of entities by size and source at the end of quarters in 2018}
\label{tab-cr-data-for-model}
\begin{tabular}{lrrrrrr}
\hline
Size & CBOP & Pracuj & Q1 & Q2 & Q3 & Q4\\
\hline
Small \& medium-sized & Yes & Yes & 100 & 129 & 107 & 76\\
                & Yes & No & 8,900 & 8,571 & 8,199 & 4,019\\
                 & No & Yes & 3,641 & 3,543 & 3,116 & 2,795\\
\hline
\multicolumn{3}{c}{$\sum$} & 12,641 & 12,243 & 11,422 & 6,890\\
\multicolumn{3}{c}{$\hat{c}$} &  0.0111 & 0.0148 & 0.0129 & 0.0186\\
\hline
Large & Yes & Yes & 534 & 582 & 552 & 303\\
     & Yes & No & 2,584 & 2,705 & 2,657 & 1,528\\
     & No & Yes & 3,780 & 3,608 & 3,506 & 3,202\\
\hline
\multicolumn{3}{c}{$\sum$} & 6,898 & 6,895 & 6,715 & 5,033\\
\multicolumn{3}{c}{$\hat{c}$} & 0.1713 & 0.1771 & 0.1720 & 0.1655\\
\hline
\end{tabular}
\end{table}

From a practical point of view and considering the possibility of using these sources in order to produce official statistics, the relationship within the groups is constant over time. The estimated value of $\hat{c}$ is close to 17\% and 2\% for the whole of 2018. This suggests that the estimated population sizes should be similar. We verified other stratification variables and concluded that the proposed grouping (large vs. other) yields the most reliable estimates in comparison with the DL survey and ensures stability over time. For more results see Appendix \ref{appen-variables}.

\subsection{Estimation results}

Table~\ref{tab-results} presents results of the estimation process. We report $\hat{N}_{\text{na\"ive}},\hat{N}_{\text{na\"ive},A},\hat{N}_{\text{na\"ive},B}$ (total, A~for small \& medium-sized firms, B for large ones) using the na\"ive estimator~\eqref{cr-petersen}, $\hat{N}_{\text{DL}},\hat{N}_{\text{DL},A},\hat{N}_{\text{DL},B}$ which is based on the DL survey, and three results obtained using model 2~\eqref{model2-ll} -- $\hat{N}_{\text{dep}},\hat{N}_{\text{dep,A}},\hat{N}_{\text{dep,B}}$, which were calculated as expected values from 500 bootstrap samples, as described in section \ref{sec-variance-est}. We calculated $\hat{N}_{\text{na\"ive}}$ without stratifying by company size; $\hat{N}_{\text{na\"ive},A},\hat{N}_{\text{na\"ive},B}$ are calculated separately by size. That is why, $\hat{N}_{\text{na\"ive},A}$ and $\hat{N}_{\text{na\"ive},B}$ do not sum up to $\hat{N}_{\text{na\"ive}}$. We decided not to estimate standard errors for $\hat{N}_{\text{na\"ive}}$ as it is highly biased and we use it only for illustrative purposes. For $\hat{N}_{\text{DL}}$ we obtained standard errors by applying the linearization method used by Statistics Poland and implemented in the \texttt{survey} package \citep{r-survey}. Finally, we report estimated parameters for the proposed model i.e. $\hat{\alpha}, \hat{p}_1, \hat{p}_{2a}$, and $\hat{p}_{2b}$.  

\begin{table}
    \centering
    \caption{The estimated number of entities with job vacancies in 2018}
    \label{tab-results}
\begin{tabular}{lrrrr}
\hline
Parameter & Q1 & Q2 & Q3 & Q4\\
\hline
\multicolumn{5}{c}{Point and standard errors estimates}\\
\hline
$\hat{N}_{\text{DL}}$ & 54,655 (2,174) &   56,076 (2,179) &  50,805  (2,031) &   44,108 (1,869) \\
$\hat{N}_{\text{na\"ive}}$ & 153,960 (--) & 132,548 (--) & 127,224 (--) & 99,694 (--)\\
$\hat{N}_{\text{dep}}$ & 73,536 (3,358) & 63,364 (2,893) & 63,724  (2,969) &  49,147 (3,278)\\
\hline
$\hat{N}_{\text{DL},A}$ & 47,393 (2,118) &   46,979 (2,037) &  42,297 (1,922) &   36,210 (1,759)  \\
$\hat{N}_{\text{na\"ive},A}$ & 336,690 (--) & 247,647 (--) & 250,189 (--) & 154,694 (--)\\
$\hat{N}_{dep,A}$ & 54,620 (2,626) & 46,004 (2,224) & 45,971 (2,275) & 33,967 (2,413)\\
\hline
$\hat{N}_{\text{DL},B}$ & 7,262 (500) &   9,097 (783) &  8,508 (669) &   7,898 (641)  \\
$\hat{N}_{\text{na\"ive},B}$ & 25,189 (--) & 23,664 (--) & 23,591 (--) & 21,180 (--)\\
$\hat{N}_{dep,B}$ & 18,916 (777) & 17,360 (707) & 17,753 (734) & 15,181 (909)\\
\hline
\multicolumn{5}{c}{95\% confidence intervals}\\
\hline
$\hat{N}_{\text{DL}}$ & 50,394; 58,916 &  51,805; 60,347 &  46,822; 54,788 &   40,443; 47,773 \\
$\hat{N}_{\text{dep}}$ & 66,866; 80,206 &  57,621; 69,107 & 57,825; 69,623 &  42,636; 55,658 \\
\hline
$\hat{N}_{\text{DL},A}$ & 43,241; 51,545 &   42,986; 50,972  &  38,530; 46,064 &   32,762; 39,658  \\
$\hat{N}_{dep,A}$  & 49,473; 59,767  & 41,646; 50,362 & 41,512; 50,431  & 29,237; 38,696 \\
\hline
$\hat{N}_{\text{DL},B}$ & 6,283; 8,241 &   7,563; 10,631 &  7,197; 9,819 &   6,641; 9,155\\
$\hat{N}_{dep,B}$ & 17,393; 20,439 & 15,975; 18,746  & 16,313; 19,192  & 13,398; 16,963\\
\hline
\multicolumn{5}{c}{Model parameters point and standard errors}\\
\hline
$\hat{\alpha}$ & 0.0690 (0.0053) & 0.0806 (0.0065) & 0.0702 (0.0058) & 0.0757 (0.0088)\\
$\hat{p}_1$ & 0.1651 (0.0077) & 0.1895 (0.0088) & 0.1810 (0.0086) & 0.1212 (0.0082)\\
$\hat{p}_{2a}$ & 0.0119 (0.0012) & 0.0161 (0.0014) & 0.0138 (0.0013) & 0.0200 (0.0023)\\
$\hat{p}_{2b}$ & 0.1837 (0.0079) & 0.1923 (0.0083)& 0.1847 (0.0080)& 0.1790 (0.0102)\\
\hline
\end{tabular}
\end{table}

First, as expected, the na\"ive estimator provides significantly higher estimates, because the number of units present in all sources is very small. This estimate would suggest that a fourth of companies in the sampling frame have a~job vacancy. The proposed estimator ($\hat{N}_{\text{dep}}$) provides a significantly lower estimate, which is closer to the estimates from the DL survey. Note that a direct comparison with the DL survey is limited as the survey covers entities and their local units and also suffers from non-response. The survey-based, estimated number of companies that reported job vacancies to PEOs differs from CBOP data (see Tables~\ref{tab-dl-survey} and~\ref{tab-statistics-sources}).

Secondly, estimates obtained from the proposed model indicate that around 10-20\% more entities reported job vacancies than in the DL survey. This result is supported by the data presented in the Appendix in Table \ref{appen-tab-statistics-sources-sizes}, where the number of large and medium-sized entities as well as the number of vacancies reported to PEOs are significantly underestimated by the DL survey. In contrast, the number of small units is overestimated. This may be due to the large rate of non-response in this group (over 60\%) and the overcoverage error of about 2\%, which resulted in large final weights assigned to these units. A comparison of confidence intevals calculated for the DL survey and our approach suggests highest discrepancies for large units, while the estimates for medium-sized and small firms are more similar. However, one should take into account that we merged local units with their central units, so this comparison may not be entirely reliable. 

The neagative dependence model is supported by the $\hat{\alpha}$ parameter,which is estimated to be around 6-8\%. It represents the proportion of behaviourally dependent individuals in the population, which, in this case, amounts to about 40-60 thousand. 

\section{Conclusions}\label{sec-concl}

In previous approaches that relied on data from online job boards the goal was to extract information about job vacancies from online sources in order to supplement data collected from probability-based surveys. This additional information can include required skills or more detailed elements of job description. Those approaches have compared distributions of online data with that observed in survey data to ensure representativeness of the former. Our aim was to provide estimates of the number of economic entities with at least one job vacancy without relying on the business sample survey. At the same time, various studies take advantage of online repositories and web-scraping to show skills demand (see e.g. \citet{colombo2018applying, deming2018skill}). In this article we describe a procedure that can be applied with a view to using such data as a source of statistics about job vacancies. The precedure consists of data collection and preparation, as well as the development of an estimator of the population of economic entities with job vacancies. We showed that job vacancy statistics can benefit from online and administrative sources, which not only provide additional variables, but also serve as the basis for estimating the total number of vacancies.

The advantage of using administrative and online data is the rich store of information provided by these sources and the lower cost of data collection in comparison with surveys. The biggest challenge encountered at the initial stages of analysis is associated with various data structures that need to be processed. During the data preparation stage it is crucial to correctly identify job titles and company names. This proves difficult in practice because of differences between administrative and online sources. The former ones rely on institutional standards (e.g. job offer codes, official classifications), while the latter are more market-oriented (e.g. are more likely to reflect emerging market trends in skill or occupational terminology). This is why, we put much effort in data preparation and linkage.

We linked data from administrative records with online data and used dual system estimation in order to estimate the number of economic entities with job vacancies in Poland. We proposed a new capture-recapture estimator, built on the approach proposed by \citet{Chatterjee2019a}, which accounts for negatively correlated sources. This approach enabled us to provide estimates of the number of entities with job vacancies based solely on non-probability sources. Additionaly, with our approach, we identified the level of bias due to non-response and under-reporting errors in the DL survey. 

Our results suggest that the DL survey underestimates the number of economic entities with at least one job vacancy by 10-20\%. The results  differ significantly depending on the size of economic entity. The number of medium-sized and large companies with vacancies is underestimated in the survey, while the number of small units is overestimated. Our simulation studies show that the proposed model
provides unbiased estimates for all parameters.

Since the methodology of the job vacancy survey in the EU is similar across member states, our approach can be applied in other countries. Moreover, online job boards in different countries rely on a similar technology, and at least some countries have detailed administrative data sources on job vacancies (see e.g. \citet{bhuller2019broadband} for Norway).

In our study we also carefully checked the collected job offers in terms of the definition of a job vacancy and found that not every online job board can be used to provide statistics about vacancies. Some advertisements do not refer to contract-based jobs or otherwise fail to  meet the definition of a job, such as B2B or contracts for specific work. The number of such offers is potentially large. However, such contracts may have a large impact on the modern labor market at the time of increasing work flexibility. In our opinion, it is necessary to reconsider the definition of a job vacancy and check to what extent the inclusion of non-standard contracts changes estimate of labor demand. Online and administrative data can be valuable sources of such information. 


\bibliographystyle{agsm.bst}

\bibliography{bibl}

\clearpage

\begin{center}
    \Large Appendix / Supplementary Materials
\end{center}

\appendix 

\renewcommand{\theequation}{\thesection.\arabic{equation}}

\section{Definitions}\label{appen-defs}

Definitions used in the Demand for Labor survey in Poland

\begin{itemize}
    \item \textbf{Reporting unit} (abbreviated form: unit) -- an entity of the national economy or its local unit, from which data are collected.
    \item \textbf{Local unit} -- An organized entity (an enterprise, a~division, a~branch, etc.) located in the place identified by a~separate address, at which or from which the activity is managed by at least one working person. In the process of identifying local units the folllowing assumptions are adopted in the National Official Business Register REGON\footnote{See \url{https://stat.gov.pl/en/metainformation/glossary/terms-used-in-official-statistics/818,term.html}}:

\begin{itemize}
    \item enterprises keeping their own accounts located at one address are treated as separate local units,
    \item  organizational parts of entities registered separately are treated as local units,
    \item  entities conducting transportation activities are treated as local units based in the place from which orders are given or where their work is organized,
    \item  entities conducting construction activities are treated as local units based in the place where construction contracts are signed, or the company's management regularly works, or where construction and auxiliary works are organized. Construction sites are not treated as local entities,
    \item  units of healthcare entities referred to in the Medical Activity Act are treated as separate local units.
\end{itemize}
\end{itemize}

\clearpage

\section{Data sources -- details, cleaning, linkage}

\subsection{Vacancy statistics in Europe}

Job vacancy statistics provide quarterly information about the level of unrealised demand for labor in all European Union (EU) member states. The data are broken down by section of economic activity, according to NACE Rev. 2 classification. Additionally, regional (NUTS) and occupational (ISCO) breakdowns are also available. However, national statistical institues are not obliged to deliver these data, which practically means that most member states do not supply this information.

Data for all countries are available from 2010Q1. Like Poland, most countries use a~sample survey for data collection. Four countries base their statistics solely on administrative sources. Data for Czechia, Croatia (since 2016) and Luxembourg come from labor offices, and in the case of Belgium -- partially from the National Social Security Office. Data from labor offices include only registered job offers and may not cover all vacancies.

Most countries treat enterprises as statistical units. Poland is one of eight exceptions in that it defines a reporting unit as an entity of the national economy or its local unit (see Appendix~\ref{appen-defs} for a~detailed definition of a~local unit). Since 2010 the survey methodology used by many countries has been revised to include companies with at least one employee but a few countries continued to exclude microcompanies for a certain period, so not every country has collected data about all enterprises for the whole period since 2010. For example, in Malta entities employing more than 1 person have been included in the job vacancy survey only since 2017.

In terms of NACE coverage, the Polish survey includes NACE sections from A to S. Quite a~few countries do not survey companies in the sector of agriculture, forestry and fishing (section A), while others exlude certain parts of service sectors. There are various data accuracy issues  in some countries. For example, in Belgium, job vacancies are believed to underrepresented in sections B and R.

There are large differences between countries regarding dates, because, as Eurostat points out, there is no international standard for recording job vacancies. The most common way is to provide the number of job vacancies at the last day of a~quarter (12 countries). Some countries report data for the middle of the quarter, as an average from three months, or provide information about the flow of vacancies throughout the quarter. Statistics Poland includes the number of vacancies at the end of a~quarter (stock of vacancies), the number of newly created jobs and the number of eliminated jobs in the quarter (flow of vacancies, but without the ISCO breakdown) and the number of newly created jobs in a~given quarter at its end (stock of vacancies). Both the stock and flow of vacancies include valuable information. The stock does not include vacancies that appeared but were filled within a~given quarter, and the flow does not include vacancies from previous periods.

\subsection{Public employment offices}\label{peo-details}

Public employment offices operate by virtue of an ordinance issued in 2014 by the Minister of Labor and Social Policy, which sets out conditions of implementing and methods of conducting labor market services (Dz.U. 2014 poz. 667). A~company interested in posting a~job offer completes a~form containing 124 fields. The form can be filled personally at a PEO, by phone or online. In the first two cases, a~PEO employee conducting an interview with the company representative determines which  classification codes should be used. If the form is filled online, it is not verified unless it looks suspicious (based on a subjective assessment). For example, when a~single job offer contains an unusually large number of vacancies (workplaces) to be filled. In such cases, a~PEO employee may call and ask for details. Interestingly, our interview with a~PEO employee from a~town with 60 thousand inhabitants revealed that about 95\% job offers are submitted either personally or by phone.

CBOP contains domestic job offers registered in PEOs, as well as foreign job offers from the EURES website maintained by the European Commission, collected by REOs, and job offers for younger persons, obtained from the VLC. We are only interested in job offers that refer to workplaces available in Poland. The majority of such offers come from PEOs. According to the ordinance that regulates the functioning of PEOs,  every domestic job offer reported to a PEO should contain the following information:

\begin{enumerate}
    \item employer data: name, address, telephone number, tax identification number, location, information whether the employer is a temporary work agency;
    \item terms of employment: location and name of the workplace, the number of vacancies, a general description of responsibilities, type of contract (15 types according to Polish Labor law), type and level of remuneration type, start date, weekly working hours, information whether the job is temporary; 
    \item job requirements: type and level of education, skills, qualifications, required languages with levels, work experience, whether foreign job seekers can apply for the position;
    \item additional information: date of job offer registration, expiration date, frequency of contacts with the employer or an employee responsible for contacts regarding the job offer, the number of workplaces to be filled, is it a~foreign job offer, an internship or an offer for a~disabled person.
\end{enumerate}

A~company can also provide information including:

\begin{enumerate}
    \item employer data: NACE section representing the company's main type of activity, preferred form of contact;
    \item terms of employment: occupation (according to the International Standard Classification of Occupations, (ISCO), but the Polish Classification of Occupations and Specialties (PCOS) is more detailed and contains 6-digit codes, i.e. two more than the ISCO classification or  the European Skills, Competences, Qualifications and Occupations (ESCO)).
\end{enumerate}

If the employer has not indicated the NACE section or the ISCO occupation code, a PEO employee is obliged to supplement this information on the basis of the company's tax identification number and the provided job description. The job offer is posted a day after its registation at the latest. It can be posted with open or restricted access. The former means that it contains publicly available information about the company, while in the latter case this information is not presented. After accepting a given job offer for processing, PEO staff check whether any unemployed person registered in this office meets the criteria specified in the job offer. In the absence of suitable candidates the office immediately informs the employer about it and notifies other PEOs about this fact using an online database of job offers.

Information about a~job offer is directed to an internal PEO system called Syriusz. This prevents the acceptance of data that do not meet certain requirements, for example offers with inappropriate PCOS codes, which should contain six digits. Nonetheless, it is possible that some detailed or sensitive data are entered incorrectly or happen to be outliers. These are usually wages and the working hours. This information may be entered in a~wrong format (fraction of weekly hours instead of the number of hours) or be simply untrue (wages). Another important detail is the date when the vacancy opens and closes. While the former is not usually problematic, the latter often is. 

The CBOP database can be used by registered entities. It can be downloaded on a~daily basis by means of a public API provided that an entity is granted access by the Polish Ministry of Development, Labor and Technology\footnote{Detailed instruction on how to download the data and a database description can be found on \url{http://oferty.praca.gov.pl/portal/instrukcja_pobierania_danych_z_cbop.pdf}}.

During a basic exploratory analysis of raw CBOP data at the end of the quarter(s), we found that employers whose job offers were published in 2018 were evenly distributed in terms of legal type (natural vs. legal person). According to Polish legislation, a sole proprietorship is a type of enterprise owned and run by one natural person, but it can also employ other people. In our database the most frequent types of companies with legal personality  were private and public limited companies. The least numerous category included public administration units, such as public kindergartens, schools and universities, municipalities, courts, research and development institutions. The least numerous group of entities with legal personality that are not part of public administration  included cooperatives, foundations, associations, private schools etc.

The biggest advantage of CBOP (compared with other sources of online job offers) is that PEO staff usually manually classify job titles according to the ISCO-08 occupational classification at the lowest level of specific occupations. The distribution of job offers from CBOP across ISCO codes (aggregated into major 1-digit groups) differed from that observed in the DL survey. Using a structural index (see $V_1$ index in \citet{jackman_roper_1987}) we estimated the mismatch index as being equal to 0.22, which means that 78\% of both distributions were similar to each other. Comparing differences between distributions across major occupational groups we found higher shares of offers in the following major occupational groups in CBOP: ``Service and sales workers'', ``Elementary occupations'', ``Technicians and associate professionals'' and ``Clerical support workers''. In the DL survey there were relatively offers more in the following occupational groups: ``Craft and related trades workers'' ``Professionals'', ``Plant and machine operators and assemblers'' and ``Managers''.

In PEOs employers more often sought low-qualified workers than professionals with advanced skills. Taking this into account, it is not sufficient to make any statistical inferences about the population of job vacancies based only on job offers obtained from CBOP.

Another distinguishing feature of job offers from CBOP was the fact that they were usually poorly described, especially as regards job responsibilities and requirements. Sometimes employers did not even provide information about required qualifications and skills. Additionally, job descriptions were usually unstructured.

\subsection{Online job boards}\label{online-selection-details}

To supplement data about vacancies we considered country-wide online recruitment services with job offers for professionals. We avoided local and employers' websites, because to capture job offers for the whole country we would need to scrape data from a~large number of websites. For the same reason we excluded Internet forums. Data from occupation-specific websites (e.g. nofluffjobs.com for IT occupations) would not be representative for a given industry. Job aggregators contain job offers from various websites but of varying quality, which it would be difficult to evaluate.

Since our focus was on country-wide online services, the question was which websites to use. According to information obtained from artefakt.pl, about 97.3\% of Internet users in Poland use the Google search engine. In addition to 
\citet{megapanel}, we used Google Trends to find the most popular Internet websites with job offers. We searched for ``praca'' (the Polish word for a~"job") to examine Google results. Excluding non-country-wide websites, the moast frequently used website turned out to be \url{https://www.pracuj.pl/} (Pracuj).

\subsubsection{Pracuj}\label{pracuj-preparation}

Pracuj is owned by a digital recruitment agency "The Pracuj Group". It operates in Poland and Ukraine. According to the company's own data, in March 2018 their website was visited by 3.1 million users.

Pracuj.pl is a job board that offers services for a fee. The cost of running a~job advertisement for a~30 days ranges from PLN 499 PLN (ca. 130 USD) up to PLN 999 PLN (ca. 260 USD). For a higher fee, it is possible to post ads for a period of 60 days. When the posting period ends, ads are classified as expired and are archived. At an extra cost, job offers can also be customised by adding a~logo or published anonymously.

\begin{figure}[ht!]
 \centering
    \includegraphics[width=0.5\textwidth]{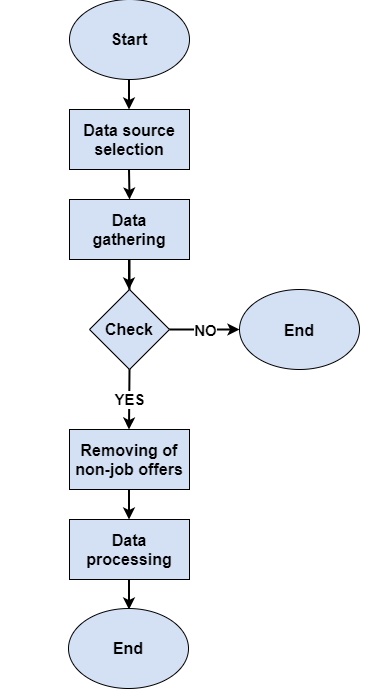}
    \caption{Data preparation algorithm}
     \label{fig:scheme}
\end{figure}

Figure~\ref{fig:scheme} presents a diagram of how job offers used in the study were prepared. After choosing the source, the data were collected, which involved downloading entire job offers as HTML documents rather than scraping information directly from the website. The number of job offers collected in 2018 is presented in Table~\ref{tab-cbop-quality} (first row).

The next step was to extract plain text from the offers. Text content was usually located in different places on the webpage, so in order to extract it we needed to identify these places. This could be done either by using the Chrome extension  \textit{SelectorGadget} or extracted directly from the HTML code. The disadvantage of Pracuj is that the same information for different job offers can be located in different places. Basic information about a~vacancy, such as job title, company name, location and dates, is usually located in the same place, but the job description is frequently included within different "div" tags. The web scraping algorithm needed to be prepared for a~certain website type.

After we identified key elements/tags of a~HTML page, we needed to extract crucial information about job vacancies from these tags. We parsed texts of all job offers to convert into a ``data table'' format. The crucial features included job title, employer (company) name, location (country, NUTS regions, city), publication and expiration dates, contract type, position level, and job description. Since HTML files contained plain text, all extracted features were treated as character vectors, even if the text included numeric, logical data and/or dates. They were converted into the appropriate format later on.

Because Polish Statistics provides quarterly estimates of labor demand, we calculated the number of job offers at the end of quarters to make the results comparable. We removed all job offers which did not meet the following conditions:

\begin{enumerate}
\item \textbf{Publication date} should be the last day of the quarter or earlier,
\item \textbf{Expiration date} should be the last day of the quarter or later.
\end{enumerate}

The number of job offers after removing those that did not meet the required conditions is presented in Table~\ref{tab-cbop-quality} (second row). This procedure helped us to significantly reduce the amount of data that had to be analyzed.

As shown in Figure~\ref{fig:scheme}, the next step involved checking whether collected job offers had been extracted properly. For example, all job offers needed to have publication and expiration dates. We made sure that jobs were located in Poland, that each location was appropriately described, etc. If at least one value in a given offer was missing, the offer was removed. In contrast to CBOP, in Pracuj we identified a~small number of job offers with wrong dates. Job offers with missing or inappropriate values (e.g. cyrillic, bad encoding) in the company name and the job title were also removed, as were offers placed by foreign employers (even those located in Poland) without a fiscal identifier (NIP or REGON). In Pracuj there were also job offers with hidden company names (i.e. \textit{hidden recruitment/anonymous advertisement}). Since this made it impossible to identify the employer such offers were also removed. The number of job offers which were removed is presented in Table~\ref{tab-cbop-quality} (third row).

After removing incomplete job offers, we removed all those that could not be classified as employment offers. We also removed all offers of jobs outside Poland, regardless of whether the employer was domestic or foreign. Next, we removed offers of traineeship, internship and voluntary work.

The biggest problem we faced at this stage was related to contract type. For some reason, job offers which had expired (even a~few months earlier) did not contain any information about contract type, unlike active ones (not expired), which contained such information. To identify contract type we analysed job descriptions. Using regular expressions we checked whether job descriptions contained types of contracts described in Section~\ref{sec-data}. If a~job offer contained information about an employment and non-employment contract, it was classified as meeting the criteria of a vacancy. It is worth noting that  employers frequently offer different types of contracts for employees to choose from. Finally, we removed job offers containing contracts for a~specific work, contracts of mandate and B2B contracts, if there was also no mention of employment contract in the job description. The number of non-job offers removed at this stage is shown in Table~\ref{tab-cbop-quality} (fifth row).

After removing non-job offers the final set of vacancies was ready for processing.  Since the job offers had been composed by different people, they contained natural language, which is characterized by much redundancy. To make job offers more comparable across the databases (CBOP and Pracuj), the following steps were applied to all character strings within both CBOP and Pracuj databases:

\begin{enumerate}
\item Convert character vectors into lowercase format,
\item Remove unusual symbols caused by bad encoding,
\item Remove legal form out of company names,
\item Remove HTML tags,
\item Remove special symbols,
\item Remove unusual spaces,
\item Remove digits from job titles,
\item Remove stop words (including casual stop words and unusual ones, e.g. contract number, place of work),
\item Perform lemmatization of job titles,
\item Remove Polish letters with diacritics.
\end{enumerate}

First of all, we converted character vectors (regions, company names, job titles etc.) into the lowercase format. Since the R programming language is case sensitive, we needed to eliminate apparent differences between strings: for example, given two versions of one company name “NESTLE” and “Nestle”, we converted them both to “nestle”. In CBOP, among 10 most frequently occurring company names we identified 3 cases of the same company with different versions of the name resulting from the use of special symbols and different letter case.

The next problem we encountered was the result of bad encoding of Polish letters. In spite of using "UTF-8" character encoding the majority of collected job offers contained inappropriate characters. Job offers were incomparable until the problem was fixed. Here are some examples of wrongly encoded letters:

\begin{enumerate}
\item InÅ¼ynier Budowy;
\item Specjalista ds. kadr i pÅ??ac;
\item Kierownik robÃ³t;
\item MÅ??odszy Projektant Elektryczny,
\end{enumerate}

The third step of processing consisted in removing legal forms from company names. We found that within the same source of data (e.g. CBOP) some company names contain different versions of legal form (e.g. public limited company or PLC). Since Pracuj requires registration, job offers were published from a~verified account, so most of them did not contain such differences.

Since individual job features were extracted from the HTML code using regular expressions, some HTML tags (e.g. \textless h1 \textgreater ... job title \textless h1 \textgreater) and their elements were also captured. All such tags were removed, as were all special symbols and unusual spaces (e.g. double and triple spaces, white spaces etc). We also removed all numeric values from job titles, which usually (particularly in CBOP) contain information about salary/wage, number of working hours and the offer number. While some numbers in job titles may have been significant, the number of cases in which they were used inappropriately was much higher.

To reduce the size of vocabulary (the number of unique words) in the analysis of job titles and to eliminate differences between words due to declension, lemmatization was performed, which involves reducing inflectional forms of a word to one single form. We also removed some standard and custom stop words to make job titles across the databases more comparable. Job titles in CBOP usually contained a~lot of unnecessary information about location, required skills, salary/wage, number of working hours, specific job offer number etc. Job titles in the Pracuj database were more laconic and accurate. We investigated the most and the least frequently occurring words in job titles and identified 300 useless words to be removed.

The last step of data processing consisted in replacing Polish letters with diacritics present in all character features (except for company names) with their Latin variants.

After this data preparation we made an attempt to compare job offers in the two data sources. To our surprise, even after data processing there was no guarantee that all job offers would me matched. We found that some offers of potentially the same jobs had different company names. Some examples of such job offers were presented in Table~\ref{cbop_pracuj_company_names}.

\begin{table}[ht!]
    \centering
    \caption{Examples of differences between company names in CBOP and Pracuj}
    \label{cbop_pracuj_company_names}
    \begin{tabular}{ll}
    \hline
    CBOP & Pracuj \\
    \hline
    tires company debica & t c debica \\
    amhil europa & wentworth technologies \\
    tema retail pl & lc waikiki \\
    tandem bedzin & authorized car showroom \\
    terg & media expert \\
    \hline
    \end{tabular}   
\end{table}

Cases of the same company (employer) with different names can probably be explained by the fact that job offers in different data sources are entered by different people.  Unfortunately, no data processing techniques can solve this problem, so we decided to replace such discrepant company names with unique alternatives. When entering job offers PEO workers provide a~lot of information about the employer, including its tax identification number (NIP), which is unique for each company in Poland. While a company may have different names (e.g. full and short names specified in the registration application), its tax identification number is always unique. Cases when the use of NIP instead of a company name allowed us to deal with discrepant names included the following:

\begin{enumerate}
\item Abbreviations (e.g. PKO BP, PKO Bank Polski are different versions of the same company name),
\item Extensions and prefixes (e.g. Nestle and Nestle Polska. We also identified some other possible extensions - Group, Partners, Company, Corporation, International etc.),
\item Parent company and their subsidiaries (e.g. Antal International (parent company) posted job offers through its subsidiaries, such as Antal SSC, Antal Finance and Accountancy, Antal Sales and Marketing, Antal Engineering and Operations).
\end{enumerate}

The biggest problem of using NIP numbers instead of company names was that, unlike CBOP, in Pracuj such information was usually not available. As a resukt, we had to identify NIP numbers on our own. Since Pracuj requires registration, some companies provided information about themselves in their profiles. Sometimes the NIP number could be identified in the employer section of the job description. Unfortunately, only some companies included their NIP numbers in their profiles or in job offers. To find NIP numbers of the remaining companies we used the Google search engine. We used  Selenium WebDriver to prepare another web scraping algorithm that would search for NIP numbers via a Google search query, formulated as follows: “company name (including legal form) + NIP + NUTS2 region”. We scraped Google results and searched for 10-digit NIP codes on certain websites (the most trusted ones). We obtained NIP codes from the REGON system (the official register of national economic entities) and compared company names from the REGON database with those collected from Pracuj. If the names matched, we stopped searching. If there were differences, they were corrected manually. Discrepancies in company names between Pracuj and REGON can be due to the following: different order of words; company names in the REGON database frequently contain first and second names of their owners; REGON often contains some extensions and prefixes (Group, Partners, Company, Corporation, International etc); the distinction between parent or subsidiary companies. To make sure that our algorithm worked properly, we manually checked 300 job offers randomly selected from both databases (150 from CBOP and 150 from Pracuj) for each quarter of 2018. In this control sample we identified 9 job offers with wrongly assigned NIP numbers.

After we identified NIP numbers of all employers in Pracuj, we used them to remove job offers published by employers that were not included in the sampling frame of the DL survey. Final counts of job offers and employers taken into account in the estimation of the number of job vacancies is presented in Table~\ref{tab-statistics-sources}.

\subsubsection{Other job boards, an example}\label{olx-service}

In the previous subsection we described the procedure of data preparation. At the stage of selecting data sources we only chose one commercial source. At the beginning of our study we considered various data sources, for example, OLX, an online marketplace with a section dovoted to recruitment. Unlike Pracuj, OLX does not specialise in employment classified ads. The cost of posting a~job offer at OLX is much lower than at Pracuj, and basic service is free. This is why we expected to capture a~variety of employers which cannot afford to or just do not want to place ads on Pracuj. The biggest advantage of OLX was that we already had job offers obtained directly from OLX. While Pracuj is a~website for employers seeking highly-skilled candidates, OLX is mainly used for recruiting low-qualified persons. In this sense it is similar to CBOP but has a much lower share of offers placed by governmental organisations. During the stage of data preparation we identified 9.3 times more job advertisements on OLX than on Pracuj. In the end, we decided not to use OLX for the reasons we present below. While this website may be a rich source of data, they did not meet all our criteria for classifying job listings as job vacancies.

Unlike CBOP and Pracuj, job offers on OLX did not usually contain company names. Instead, names of people responsible for handling inquiries from potential candidates were given (see examples in Table~\ref{olx_company_names}). Unfortunately, these names could not be used to identify specific employers since in the case of "Piotr" we identified 94,767 significantly different job offers. We could analyse job offers on the OLX website but could not estimate the number of employers that had posted vacancies or their characteristics. Only for some job offers could company names be extracted from the job description.

\begin{table}[ht!]
    \centering
    \caption{The most frequently occurs company names across different source of data}
    \label{olx_company_names}
    \begin{tabular}{lll}
    \hline
    CBOP & Pracuj & OLX \\
    \hline
    DINO POLSKA S.A. & Jeronimo Martins Polska S.A. & Piotr \\
    POCZTA POLSKA S.A. & Hays Poland & Marcin \\
    HR SYSTEMS sp. z o.o. & Grafton Recruitment & Agnieszka\\
    \hline
    \end{tabular}   
\end{table}

The second problem we encountered with OLX job offers was the lack of information about job location. Since we were interested in job offers located in Poland, workplaces outside Poland had to be removed. We analysed some job offers where the indicated location was in Poland but found that in fact they described foreign jobs. Table~\ref{olx_location} contains some examples of foreign job offers with a~Polish location.

\begin{table}[ht!]
    \centering
    \caption{Actual vs indicated location of job offers in OLX}
    \label{olx_location}
    \begin{tabular}{lll}
    \hline
    Job title & Indicated location \\
    \hline
    Carer for the elderly in Germany  & Warsaw (Poland) \\
    Order Picker Germany Magdeburg & Otrebusy (Poland) \\
    Welder MIG or MAG Germany & Tarnowskie Góry (Poland) \\
    Forklift operator Germany Strullendorf & Poznań (Poland) \\
    \hline
    \end{tabular}   
\end{table}

The last issue involved offers which were not work contracts. Many advertisements referred to agreements for a specific single job assignment, sometimes lasting only a few days, or to other forms of work that did not meet the criteria of a work contract.  Some examples are provided below. Because many of them did not contain accurate information about the position or contract type and the description was short, we were not able to assess them.

\begin{enumerate}
\item Dam Prace Sosnowiec/Dam zlecenie (Job offer in Sosnowiec / Single work contract)
\item Przyjme zlecenie pracy dla ekipy 4-osobowej (I will accept an order contract for a 4-person team)
\item Asystent Grupy Projektowej - umowa zlecenia (Project Group Assistant - contract of mandate)
\end{enumerate}

Based on the findings presented above, we concluded that OLX did not satisfy our requirements, so we removed it from our analysis. This example shows that not all sources of job offers can be used for estimating the number of vacancies.

\subsection{Results by company size}\label{detailed-results}

Table \ref{appen-tab-statistics-sources-sizes} presents the estimated number of entities and vacancies based on the Demand for Labor survey, CBOP and Pracuj after data cleaning and truncation to 30 days. 

\begin{table}[ht!]
    \centering
     \caption{Number of entities and vacancies according to the DL survey, CBOP and Pracuj, based on truncated data}
    \label{appen-tab-statistics-sources-sizes}
    \begin{tabular}{ll|rrrr}
    \hline
     Source &  Variable  & 2018Q1 &  2018Q2 & 2018Q3 & 2018Q4  \\
    \hline
     \multicolumn{6}{c}{Entities} \\
            \hline
        DL Survey    & Large & 1,253 & 2,115 & 2,092 & 1,585 \\
            & Medium  & 3,362 & 4,398 & 3,465 & 2,920 \\ 
            & Small  & 9,962 & 8,987 & 6,162 & 5,484 \\
            \cline{2-6}
     CBOP       & Large & 3,118 & 3,287 & 3,209 & 1,831 \\
            & Medium  & 4,068 & 4,017 & 3,876 & 1,984 \\ 
            & Small  & 4,932 & 4,683 & 4,430 & 2,111 \\
             \cline{2-6}
      Pracuj      & Large & 4,314 & 4,190 & 4,058 & 3,505 \\
            & Medium  & 2,337 & 2,255 & 2,003 & 1,682 \\ 
            & Small  & 1,404 & 1,417 & 1,220  & 1,189 \\
            \hline
\multicolumn{6}{c}{Vacancies } \\      
            \hline
        DL Survey   & Large  & 7,681 & 10,524 & 10,617 & 9,033 \\ 
            & Medium  & 8,571 & 9,593 & 8,586 & 6,723 \\ 
            & Small & 18,929 & 14,015 & 11,080 & 9,581 \\ 
            \cline{2-6}
            \cline{2-6}
       CBOP     & Large  & 21,514 & 26,441 & 22,592 & 11,199 \\ 
            & Medium  & 10,634 & 9,922 & 9,457 & 4,614 \\ 
            & Small & 10,259 & 9,617 & 8,573 & 4,016 \\ 
            \cline{2-6}
            \cline{2-6}
       Pracuj     & Large  & 26,187 & 26,921 & 28,112 & 21,150 \\ 
            & Medium  & 5,918 & 5,519 & 5,254 & 3,858 \\ 
            & Small & 3,560 & 3,577 & 3,152 & 2,525 \\ 
\hline            
    \end{tabular}
\end{table}

\clearpage

\section{Methods -- Derivation of MLE} \label{app-model2-mle}





\subsection{Log-likelihood}

Log-likelihood function is given by

\begin{equation}
\begin{aligned} 
\log L\left(\mathbf{\theta} \mid \mathbf{x}_{A}, \mathbf{x}_{B}\right) =  & 
\log(N_A!) + \log(N_B!) - (\log( (N_A-x_{0A})!) + \log( (N_B-x_{0B})!))  \\
& + x_{11A}\log( (1-\alpha)p_1 p_{2A}) + x_{11B} \log((1-\alpha)p_1 p_{2B}) \\
& + x_{10A} \log(p_1(\alpha  + (1-\alpha)(1-p_{2A}))) \\
& + x_{10B} \log( p_1(\alpha + (1-\alpha)(1-p_{2B}))) \\
& + x_{01A} \log((1-p_1) (\alpha  + (1-\alpha)p_{2A})) \\ 
& + x_{01B} \log((1-p_1)(\alpha  + (1-\alpha)p_{2B})) \\
& + \left(N_{A}-x_{0 A}\right) \log((1-\alpha)(1-p_1)(1-p_{2A})) \\ 
& + \left(N_{B}-x_{0 B}\right) \log((1-\alpha)(1-p_1)(1-p_{2B})).
\end{aligned}
\end{equation}

For $\log(x!)$ we use the following approximation: $x\log(x) - x$ and the log-likelihood becomes:

\begin{equation}
\begin{aligned} 
\log L\left(\mathbf{\theta} \mid \mathbf{x}_{A}, \mathbf{x}_{B}\right) \approx & 
N_A\log(N_A) - N_A + N_B\log(N_B) - N_B \\
& - (N_A-x_{0A})\log(N_A-x_{0A}) + (N_A-x_{0A}) \\
& - (N_B-x_{0B})\log(N_B-x_{0B}) + (N_B-x_{0B}) \\
& + x_{11A}\log( (1-\alpha)p_1 p_{2A}) + x_{11B} \log((1-\alpha)p_1 p_{2B}) \\
& + x_{10A} \log(p_1(\alpha  + (1-\alpha)(1-p_{2A}))) \\
& + x_{10B} \log( p_1(\alpha + (1-\alpha)(1-p_{2B}))) \\
& + x_{01A} \log((1-p_1) (\alpha  + (1-\alpha)p_{2A})) \\ 
& + x_{01B} \log((1-p_1)(\alpha  + (1-\alpha)p_{2B})) \\
& + \left(N_{A}-x_{0 A}\right) \log((1-\alpha)(1-p_1)(1-p_{2A})) \\ 
& + \left(N_{B}-x_{0 B}\right) \log((1-\alpha)(1-p_1)(1-p_{2B})),
\end{aligned}
\label{eq-log-ll-appen}
\end{equation}

where $\mathbf{\theta} = (N_A, N_B, \alpha, p_1, p_{2A}, p_{2B})$. First derivatives of~\eqref{eq-log-ll-appen} are given by

\begin{equation}
    \frac{\partial \log L}{\partial N_A} = \log N_A - \log(N_A-x_{0A}) + \log( (1-\alpha)(1-p_1)(1-p_{2A})),
\end{equation}

\begin{equation}
    \frac{\partial \log L}{\partial N_B} = \log N_B - \log(N_B-x_{0B}) +  \log( (1-\alpha)(1-p_1)(1-p_{2B})),
\end{equation}

\begin{equation}
    \begin{aligned}
    \frac{\partial \log L}{\partial \alpha}  & = -\frac{x_{11A}}{1-\alpha} - \frac{x_{11B}}{1-\alpha} + \frac{x_{10A}p_{2A}}{\alpha + (1-\alpha)(1-p_{2A})} + \frac{x_{10B}p_{2B}}{\alpha + (1-\alpha)(1-p_{2B})} \\
    & + \frac{x_{01A}(1-p_{2A})}{\alpha + (1-\alpha)p_{2A}} + \frac{x_{01B}(1-p_{2B})}{\alpha + (1-\alpha)p_{2B}} - \frac{N_A - x_{0A}}{1-\alpha} -\frac{N_B - x_{0B}}{1-\alpha},
    \end{aligned}
\end{equation}

\begin{equation}
\begin{aligned}
    \frac{\partial \log L}{\partial p_1} & = \frac{x_{11A}}{p_1} + \frac{x_{11B}}{p_1} + \frac{x_{10A}}{p_1} + \frac{x_{10B}}{p_1} - \frac{x_{01A}}{(1-p_1)} - \frac{x_{01B}}{(1-p_1)} - \frac{N_A - x_{0A}}{1-p_1} - \frac{N_B - x_{0B}}{1-p_1} \\
    & = \frac{x_{11A} + x_{11B} + x_{10A} + x_{10B}}{p_1} - \frac{x_{01A} + x_{01B} + (N_A - x_{0A}) + (N_B - x_{0B})}{1-p_1},
\end{aligned}
\end{equation}

\begin{equation}
    \frac{\partial \log L}{\partial p_{2A}}  =  \frac{x_{11A}}{p_{2A}} - \frac{x_{10A}(1-\alpha)}{\alpha + (1-\alpha)(1-p_{2A})} + 
    \frac{x_{01A}(1-\alpha)}{\alpha + (1-\alpha)p_{2A}} - \frac{N_A-x_{0A}}{1-p_{2A}},
\end{equation}

\begin{equation}
    \frac{\partial \log L}{\partial p_{2B}} =  \frac{x_{11B}}{p_{2B}} - \frac{x_{10B}(1-\alpha)}{\alpha + (1-\alpha)(1-p_{2B})} + 
    \frac{x_{01B}(1-\alpha)}{\alpha + (1-\alpha)p_{2B}} - \frac{N_B-x_{0B}}{1-p_{2B}}.
\end{equation}

Second derivatives of~\eqref{eq-log-ll-appen} are given by


\begin{equation}
    \frac{\partial^2 \log L}{\partial N_A^2} = \frac{1}{N_A} - \frac{1}{N_A-x_{0A}}
\end{equation}

\begin{equation}
    \frac{\partial^2 \log L}{\partial N_A \partial N_B} = 0,
\end{equation}

\begin{equation}
    \frac{\partial^2 \log L}{\partial N_A \partial \alpha} = \frac{-1}{1-\alpha},
\end{equation}

\begin{equation}
    \frac{\partial^2 \log L}{\partial N_A \partial p_1} = \frac{-1}{1-p_1},
\end{equation}

\begin{equation}
    \frac{\partial^2 \log L}{\partial N_A \partial p_{2A}} = \frac{-1}{1-p_{2A}},
\end{equation}

\begin{equation}
    \frac{\partial^2 \log L}{\partial N_A \partial p_{2B}} = 0,
\end{equation}


\begin{equation}
    \frac{\partial^2 \log L}{\partial N_B \partial N_A} = 0,
\end{equation}

\begin{equation}
    \frac{\partial^2 \log L}{\partial N_B^2} = \frac{1}{N_B} - \frac{1}{N_B-x_{0B}},
\end{equation}

\begin{equation}
    \frac{\partial^2 \log L}{\partial N_B \partial \alpha} = \frac{-1}{1-\alpha},
\end{equation}

\begin{equation}
    \frac{\partial^2 \log L}{\partial N_B \partial p_1} = \frac{-1}{1-p_1},
\end{equation}

\begin{equation}
    \frac{\partial^2 \log L}{\partial N_B \partial p_{2A}} = 0,
\end{equation}

\begin{equation}
    \frac{\partial^2 \log L}{\partial N_B \partial p_{2A}} = \frac{-1}{1-p_{2B}},
\end{equation}


\begin{equation}
    \frac{\partial^2 \log L}{\partial \alpha \partial N_A} = -\frac{1}{1-\alpha},
\end{equation}

\begin{equation}
    \frac{\partial^2 \log L}{\partial \alpha \partial N_B} = -\frac{1}{1-\alpha},
\end{equation}

\begin{equation}
\begin{aligned}
    \frac{\partial^2 \log L}{\partial \alpha^2} & =
    -\frac{x_{11A}}{(1-\alpha)^2} -\frac{x_{11B}}{(1-\alpha)^2} 
    - \frac{x_{10A}p_{2A}^2}{(\alpha + (1-\alpha)(1-p_{2A}))^2} - \frac{x_{10B}p_{2B}^2}{(\alpha + (1-\alpha)(1-p_{2B}))^2}  \\
    & -\frac{x_{10A}(1-p_{2A})^2}{(\alpha + (1-\alpha)p_{2A})^2}
    - \frac{x_{10B}(1-p_{2B})^2}{(\alpha + (1-\alpha)p_{2B})^2}  - \frac{N_A-x_{0A}}{(1-\alpha)^2} - \frac{N_B-x_{0B}}{(1-\alpha)^2}
    \end{aligned}
\end{equation}

\begin{equation}
    \frac{\partial^2 \log L}{\partial \alpha \partial p_1} = 0,
\end{equation}

\begin{equation}
    \frac{\partial^2 \log L}{\partial \alpha \partial p_{2A}} = 
    \frac{x_{10A}}{(1-p_{2A}+\alpha p_{2A})^2} -
    \frac{x_{01A}}{(\alpha + (1-\alpha)p_{2A})^2},
\end{equation}

\begin{equation}
    \frac{\partial^2 \log L}{\partial \alpha \partial p_{2B}} = 
    \frac{x_{10B}}{(1-p_{2B}+\alpha p_{2B})^2} -
    \frac{x_{01B}}{(\alpha + (1-\alpha)p_{2B})^2},
\end{equation}


\begin{equation}
    \frac{\partial^2 \log L}{\partial p_1 \partial N_A} = 0,
\end{equation}

\begin{equation}
    \frac{\partial^2 \log L}{\partial p_1 \partial N_B} = 0,
\end{equation}

\begin{equation}
    \frac{\partial^2 \log L}{\partial p_1 \partial \alpha} = 0,
\end{equation}

\begin{equation}
    \frac{\partial^2 \log L}{\partial p_1^2} = -\frac{x_{11A} + x_{11B} + x_{10A} + x_{10B}}{p_1^2} - \frac{x_{01A} + x_{01B} + (N_A - x_{0A}) + (N_B - x_{0B})}{(1-p_1)^2},
\end{equation}

\begin{equation}
    \frac{\partial^2 \log L}{\partial p_1 \partial p_{2A}} = 0,
\end{equation}

\begin{equation}
    \frac{\partial^2 \log L}{\partial p_1 \partial p_{2B}} = 0,
\end{equation}


\begin{equation}
    \frac{\partial^2 \log L}{\partial p_{2A} \partial N_A} = - \frac{N_A}{1-p_{2A}},
\end{equation}

\begin{equation}
    \frac{\partial^2 \log L}{\partial p_{2A} \partial N_B} = 0,
\end{equation}

\begin{equation}
    \begin{aligned}
    \frac{\partial^2 \log L}{\partial p_{2A} \partial \alpha} & = 
    \frac{x_{10A}}{(\alpha + (1-\alpha)(1-p_{2A}))^2} -
    \frac{x_{01A}}{(\alpha - (1-\alpha)p_{2A})^2}
    \end{aligned},
\end{equation}

\begin{equation}
    \frac{\partial^2 \log L}{\partial p_{2A} \partial p_1} = 0
\end{equation}

\begin{equation}
\begin{aligned}
    \frac{\partial^2 \log L}{\partial p_{2A}^2} &= -\frac{x_{11A}}{p_{2A}^2} + \frac{x_{01A}(1-\alpha)^2}{(\alpha + (1-\alpha)(1-p_{2A}))^2} \frac{x_{01A}(1-\alpha)^2}{(\alpha + (1-\alpha)p_{2A})^2} + 
    \frac{N_A-x_{0A}}{(1-p_{2A})^2}, 
\end{aligned}
\end{equation}

\begin{equation}
    \frac{\partial^2 \log L}{\partial p_{2A} \partial p_{2B}} = 0,
\end{equation}


\begin{equation}
    \frac{\partial^2 \log L}{\partial p_{2B} \partial N_A} = 0,
\end{equation}

\begin{equation}
    \frac{\partial^2 \log L}{\partial p_{2B} \partial N_B} = - \frac{N_B}{1-p_{2B}},
\end{equation}

\begin{equation}
    \frac{\partial^2 \log L}{\partial p_{2B} \partial \alpha} = \frac{x_{10B}}{(\alpha + (1-\alpha)(1-p_{2B}))^2} -
    \frac{x_{01B}}{(\alpha - (1-\alpha)p_{2B})^2},
\end{equation}

\begin{equation}
    \frac{\partial^2 \log L}{\partial p_{2B} \partial p_1} = 0,
\end{equation}

\begin{equation}
    \frac{\partial^2 \log L}{\partial p_{2B} \partial p_{2A}} = 0,
\end{equation}

\begin{equation}
    \frac{\partial^2 \log L}{\partial p_{2B}^2} = -\frac{x_{11B}}{p_{2B}^2} + \frac{x_{01B}(1-\alpha)^2}{(\alpha + (1-\alpha)(1-p_{2B}))^2} \frac{x_{01B}(1-\alpha)^2}{(\alpha + (1-\alpha)p_{2B})^2} + 
    \frac{N_B-x_{0B}}{(1-p_{2B})^2}.
\end{equation}



\clearpage

\section{Selection of stratification variables}\label{appen-variables}

Table \ref{appen-tab-strata-vars} presents a sensitivity analysis with respect to the selection of stratification variables. Because stratification variables could only have two levels, in the case of variables with more levels, the other levels had to be aggregated (e.g. Size == “Large” and Size != “Large”).  For some NACE sections, public sector of ownership and Mazowieckie province, the number of entities is high, even over 90-100k, which is significantly higher than that reported by the DL survey. On the other hand, for size == “Medium” and NACE == “G”, the number is significantly underestimated. That is why, we decided to use size == “Large”.

\begin{table}[ht!]
    \centering
    \caption{Population sizes with different combinations of stratification variables}
    \label{appen-tab-strata-vars}
\begin{tabular}{lrrrr}
\hline
Variable and selected level & Q1 & Q2 & Q3 & Q4 \\ 
\hline
size = Small & 99,605 & 81,357 & 80,729 & 61,077\\
size = Medium & 31,141 & 31,766 & 36,105 & 33,163\\
sector = Public & 104,583 & 124,812 & 97,505 & 71,831\\
NACE = C & 141,674 & 126,785 & 121,292 & 95,381\\
NACE = G & 19,845 & 19,405 & 18,370 & 28,239\\
NACE = I & 142,199 & 122,570 & 119,051 & 93,220\\
NACE = M & 101,602 & 125,072 & 81,941 & 55,319\\
Province = Mazowieckie & 132,167 & 111,632 & 111,428 & 91,009\\
\hline
\end{tabular}
\begin{flushleft}
Note: C -- Manufacturing; G -- Trade; repair of motor vehicles; I -- Accommodation and catering; J -- Information and communication;  M -- Professional, scientific and technical activities.
\end{flushleft}
\end{table}

Wwe calculated corresponding results for all NACE sections and provinces. Because of a low number of entities in some NACE sections the algorithm did not converge for most levels. For provinces, all levels yield similar results to the one presented above. These results are only reported in the replication materials (within Jupyter Notebooks).

\section{Simulation studies}\label{sec-sim-study}

\subsection{Simulation study 1}

In the first simulation, we verified the performance of the nai\"ve and the proposed estimator using data that resemble the target population. The simulation was conducted as follows:

\begin{enumerate}
    \item we generate probabilities according to model~\eqref{model-neg} given by~\eqref{eq-model1-probs} with the following:
    \begin{itemize}
        \item $N_A=50,000$, $\alpha=0.05$, $p_1=0.15$ and $p_{2a}=0.05$,
        \item $N_A=20,000$, $\alpha=0.05$, $p_1=0.15$ and $p_{2a}=0.15$,
    \end{itemize}
    \item in each of 500 iterations:
    \begin{itemize}
        \item we independently generate counts $(x_{11}, x_{10}, x_{01}, x_{00})$ for each sub-population $A$ and $B$ from the multinomial distribution of size defined by $N_A$ and $N_B$ respectively,
        \item we use vector $(x_{11A}, x_{10A}, x_{01A},x_{11B}, x_{10B}, x_{01B})$ to obtain two estimators -- na\"ive and the proposed one -- for $N_A$ and $N_B$.
    \end{itemize}
    \item finally, we calculate the expected value, relative bias and the coefficient of variation.
\end{enumerate}

Results of this simulation study are presented in Table~\ref{app-tab-sim-1}. As expected, when the observed counts are generated according to the assumption of negative dependence, the proposed model provides unbiased estimates of all parameters. The nai\"ve estimator is significantly biased, with relative bias increasing with the growing population size. 

\begin{table}[ht!]
    \centering
    \caption{Results of simulation study 1 ($N_A=50,000$, $N_B=20,000$)}
    \label{app-tab-sim-1}
   \begin{tabular}{lrrr}
   \hline
Parameter & Expected value & Relative Bias [\%] & Coeff. of variation [in \%]\\
\hline
& \multicolumn{3}{c}{Na\"ive} \\
\hline
$N_A$ & 95,259 & 47.5113 & 4.8531\\
$N_B$ & 26,033 & 23.1732 & 4.2945\\
\hline
& \multicolumn{3}{c}{Proposed} \\
\hline
$N_A$ & 50,298 & 0.5920 & 8.0528\\
$N_B$ & 20,096 & 0.4791 & 7.3090\\
$\alpha$ & 0.0504 & 0.6958 & 20.4676\\
$p_1$ & 0.1500 & -0.0107 & 7.7650\\
$p_{2a}$ & 0.0498 & -0.3571 & 4.6520\\
$p_{2b}$ & 0.1500 & 0.0283 & 5.2153\\
\hline
\end{tabular}
\end{table}

Table~\ref{app-tab-sim-1-coverage} shows the degree of coverage provided by the confidence intervals based on estimated standard errors and the suggested interval given by~\eqref{eq-ci-proposed}. As can be seen, the confidence intervals have similar lower and upper bounds, and that their coverage, based on 10000 replications, is within the nominal 95\%.

\begin{table}[ht!]
\centering
\caption{95\% confidence intervals and its coverage based on simulation study 1}
\label{app-tab-sim-1-coverage}
\begin{tabular}{lrrr}
\hline
Parameter & Lower & Upper & Coverage \\
\hline
& \multicolumn{3}{c}{Standard} \\
\hline
$N_A$ & 42,359 & 58,236 & 0.9600\\
$N_B$ & 17,217 & 22,975 & 0.9560\\
\hline
& \multicolumn{3}{c}{\citet[p. 11]{Chatterjee2019a}} \\
\hline
$N_A$ & 43,139 & 59,083 & 0.9620\\
$N_B$ & 17,504 & 23,286 & 0.9600\\
\hline
\end{tabular}

\end{table}

\subsection{Simulation study 2}

In the second simulation study we checked how the proposed method performs when the main assumption i.e. $p_{1a}=p_{1b}=p_1$ is violated and analysed its impact onbias and the root mean square error (RMSE) of $\hat{N}_A$, $\hat{N}_B$ and $\hat{N}=\hat{N}_A+\hat{N}_B$. We assumed the same population sizes and $\alpha$ as in Simulation~1 and we considered three scenarios:

\begin{enumerate}
    \item Scenario 1:
    \begin{itemize}
        \item $p_{1a} = 0.15$,
        \item $p_{1b} = [0.01,0.02,...,0.34,0.35]$,
        \item $p_{2a} = 0.05$ and $p_{2b} = 0.15$,
    \end{itemize}
    \item Scenario 2:
        \begin{itemize}
        \item $p_{1a} = [0.01,0.02,...,0.34,0.35]$,
        \item $p_{1b} = 0.15$,
        \item $p_{2a} = 0.05$ and $p_{2b} = 0.15$,
    \end{itemize}
    \item Scenario 3:
        \begin{itemize}
        \item $p_{1a} = 0.15$,
        \item $p_{1b} = [0.01,0.02,...,0.34,0.35]$
        \item $p_{2a} = p_{2b} = 0.15$.
    \end{itemize}
\end{enumerate}

\begin{figure}[ht!]
    \centering
    \includegraphics[width=0.45\textwidth]{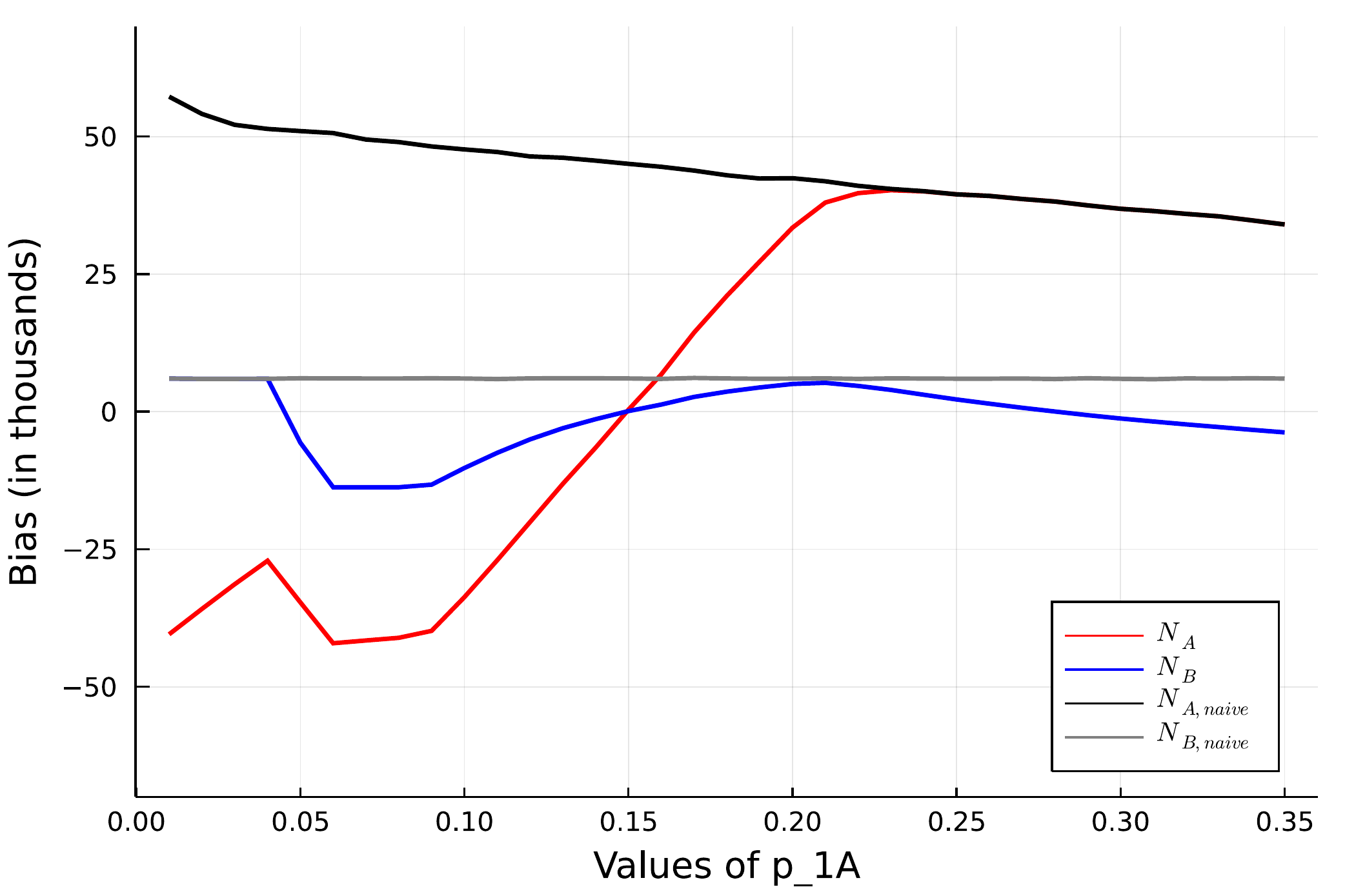}
    \includegraphics[width=0.45\textwidth]{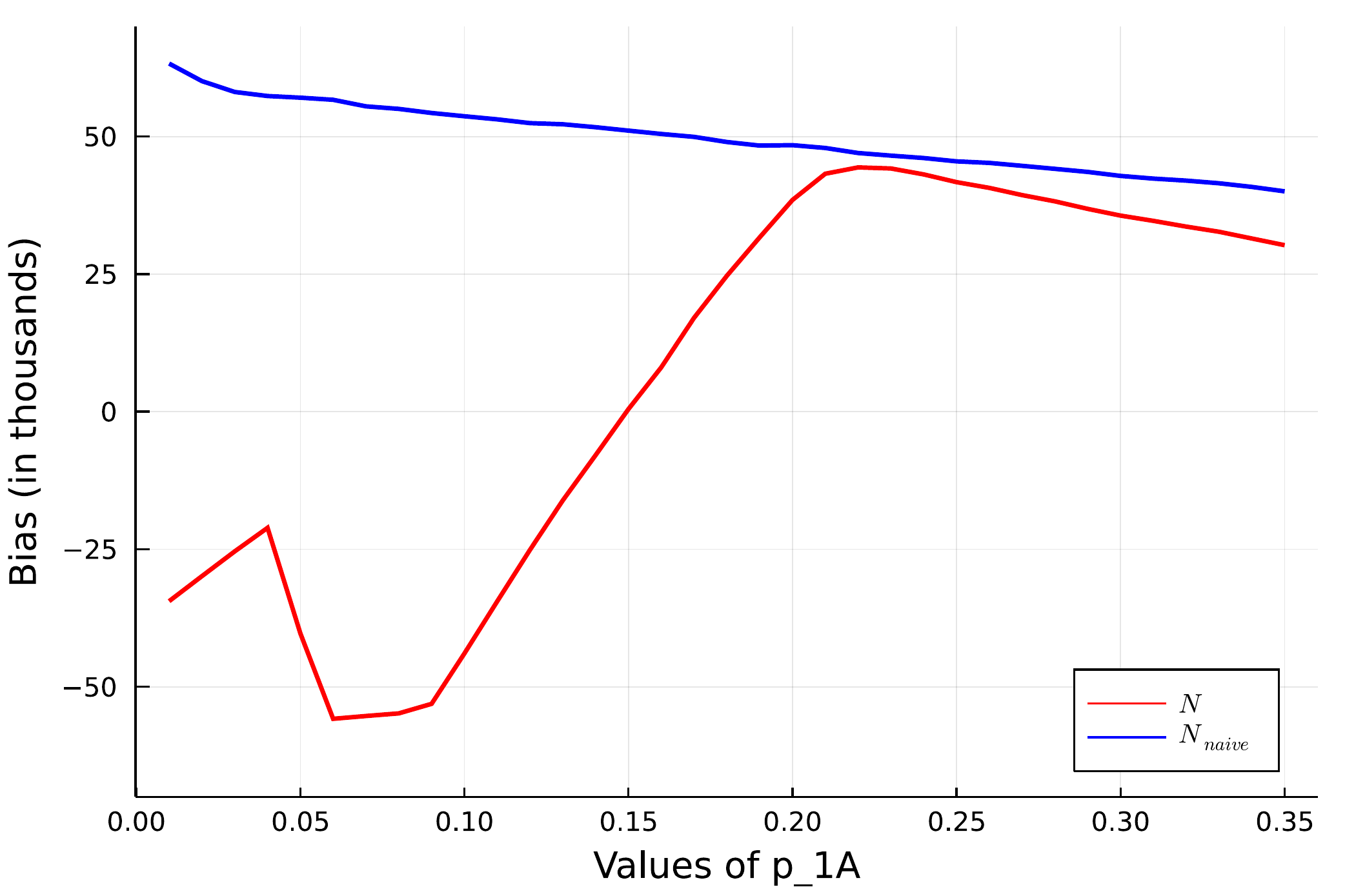}
    \includegraphics[width=0.45\textwidth]{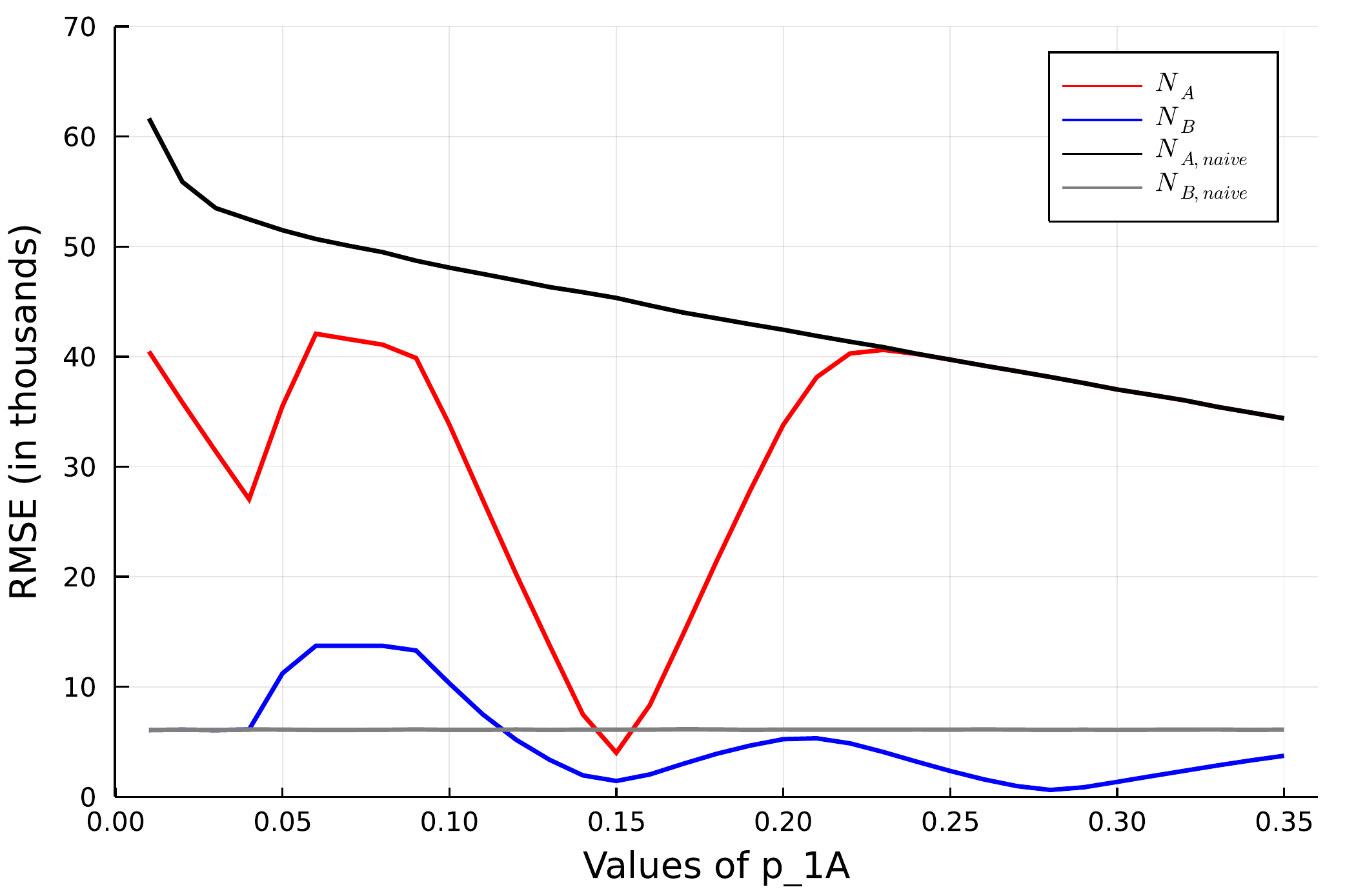}
    \includegraphics[width=0.45\textwidth]{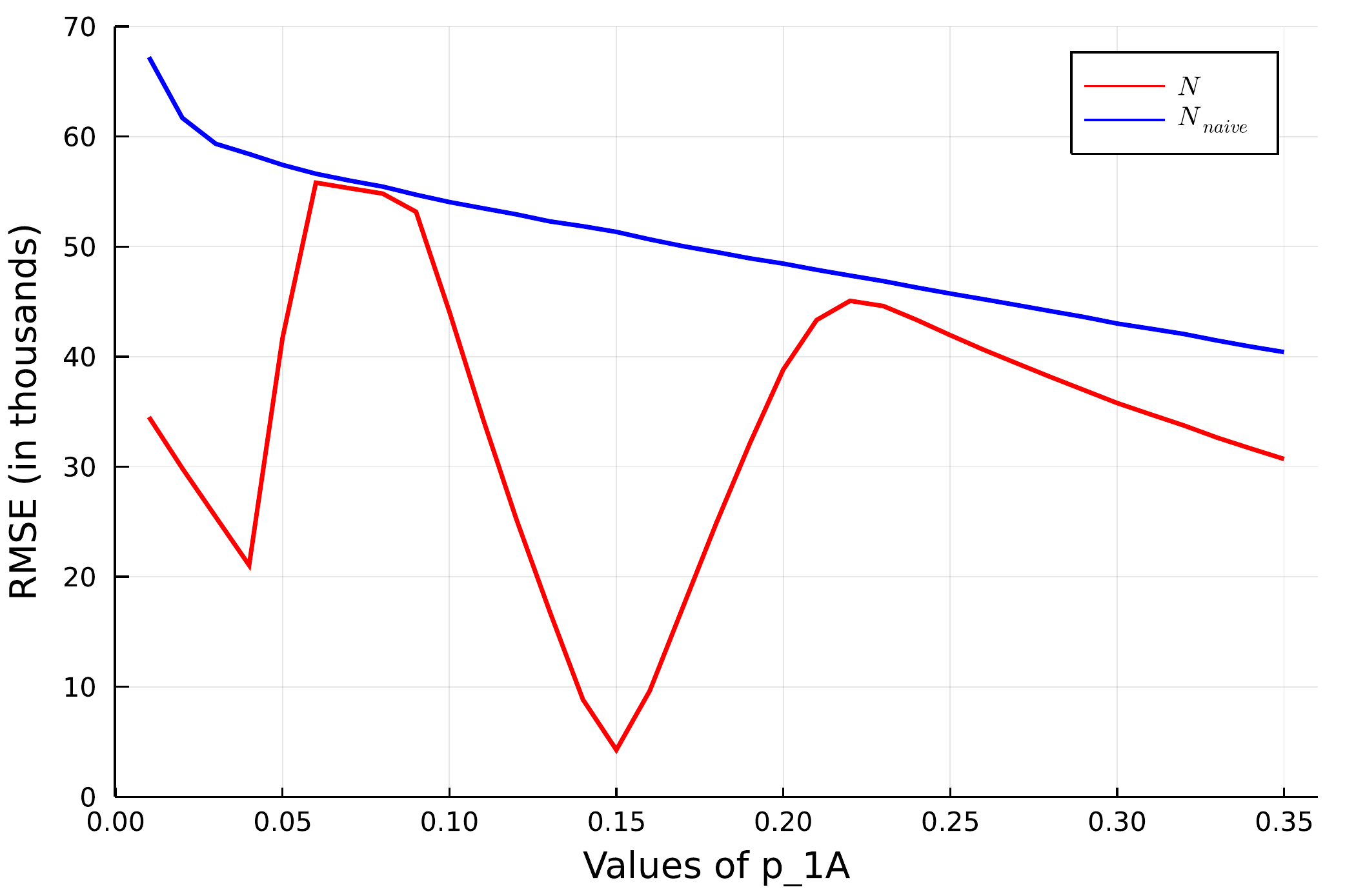}
    \caption{Bias (top) and RMSE (bottom) of the proposed estimator depending on the value of $p_{1A}$ by sub-populations (left) and the whole population (right) based on scenario 1}
    \label{appen-fig-sim2-s1}
\end{figure}

\begin{figure}[ht!]
    \centering
    \includegraphics[width=0.45\textwidth]{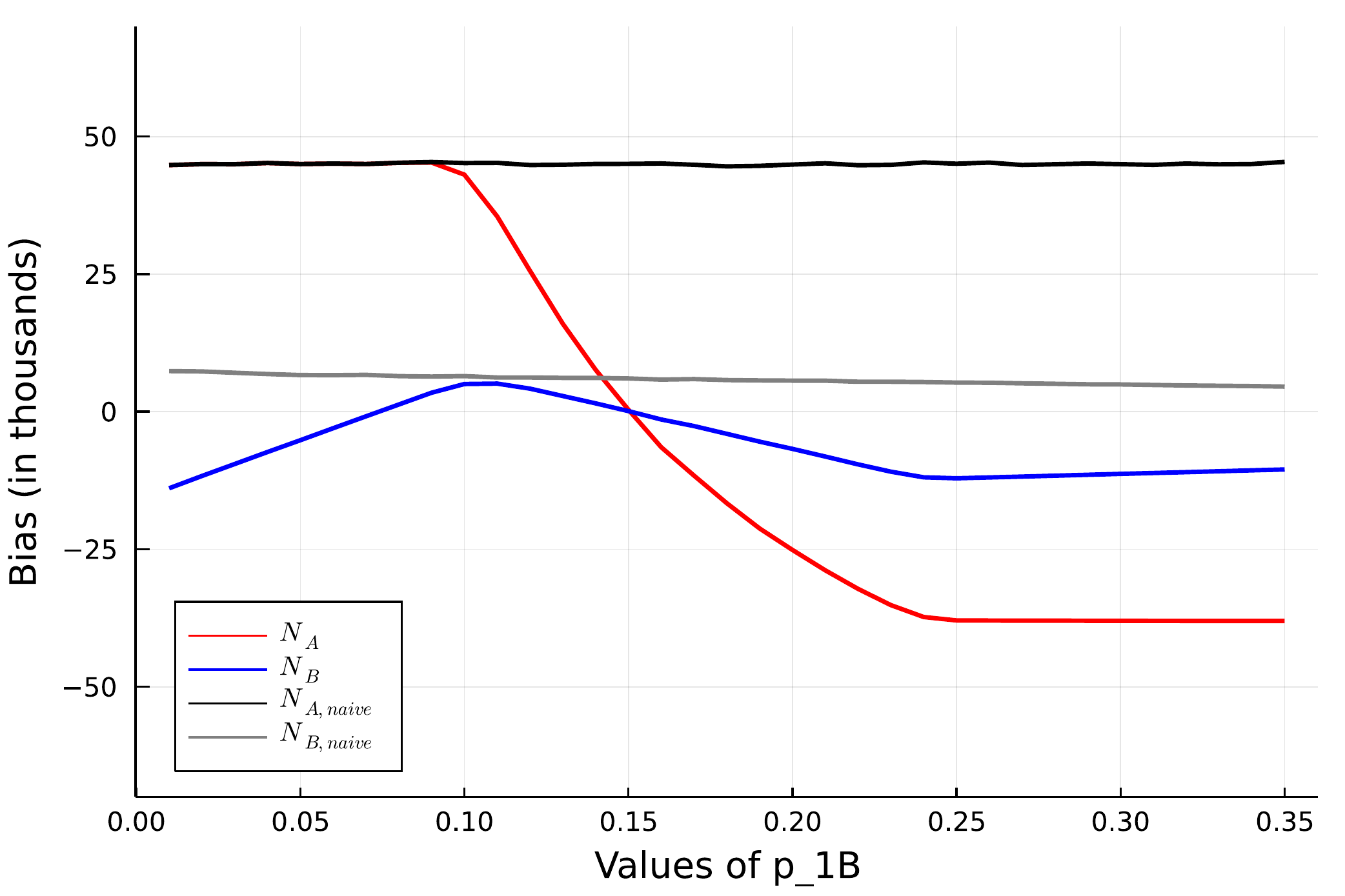}
    \includegraphics[width=0.45\textwidth]{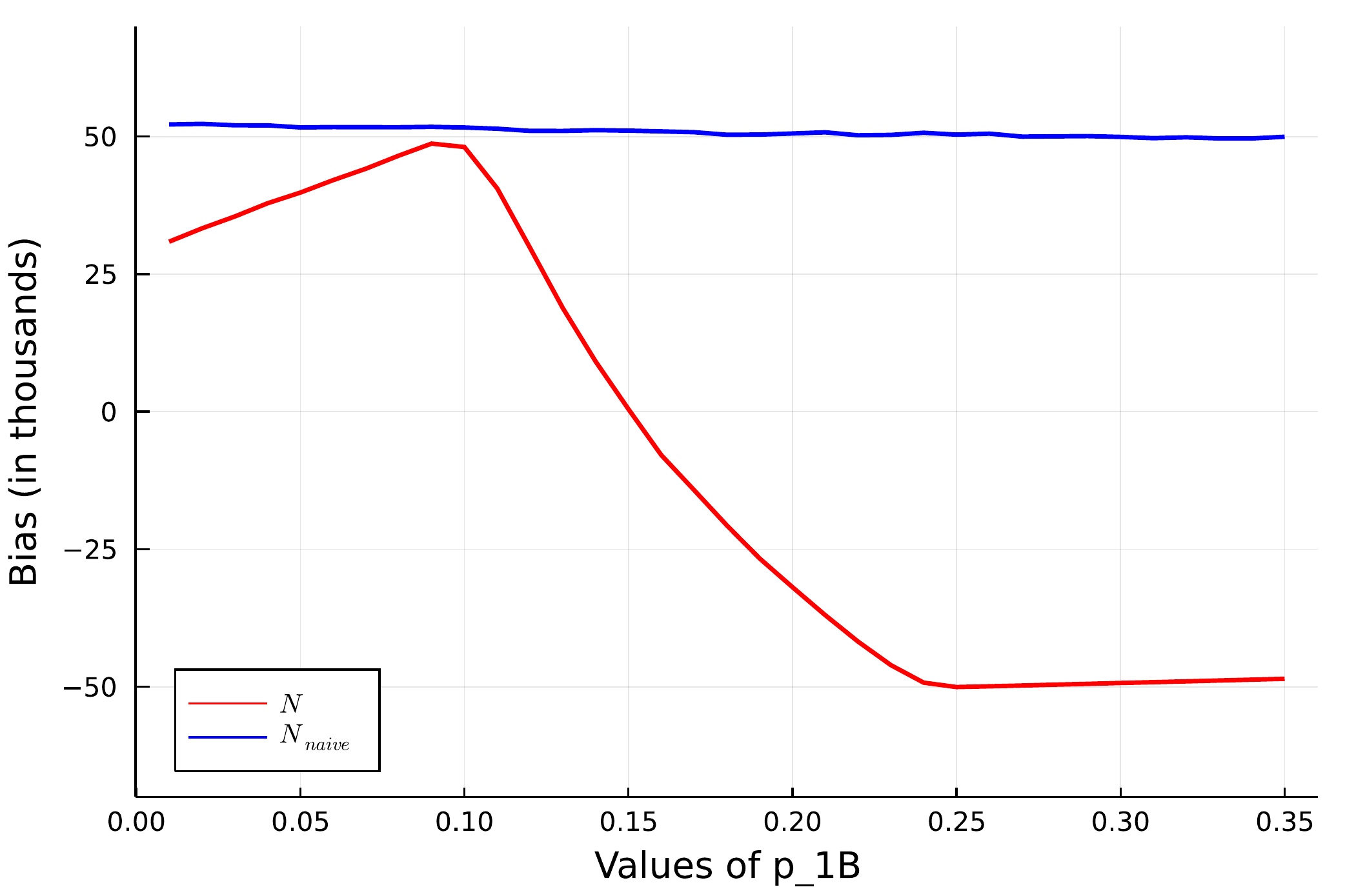}
    \includegraphics[width=0.45\textwidth]{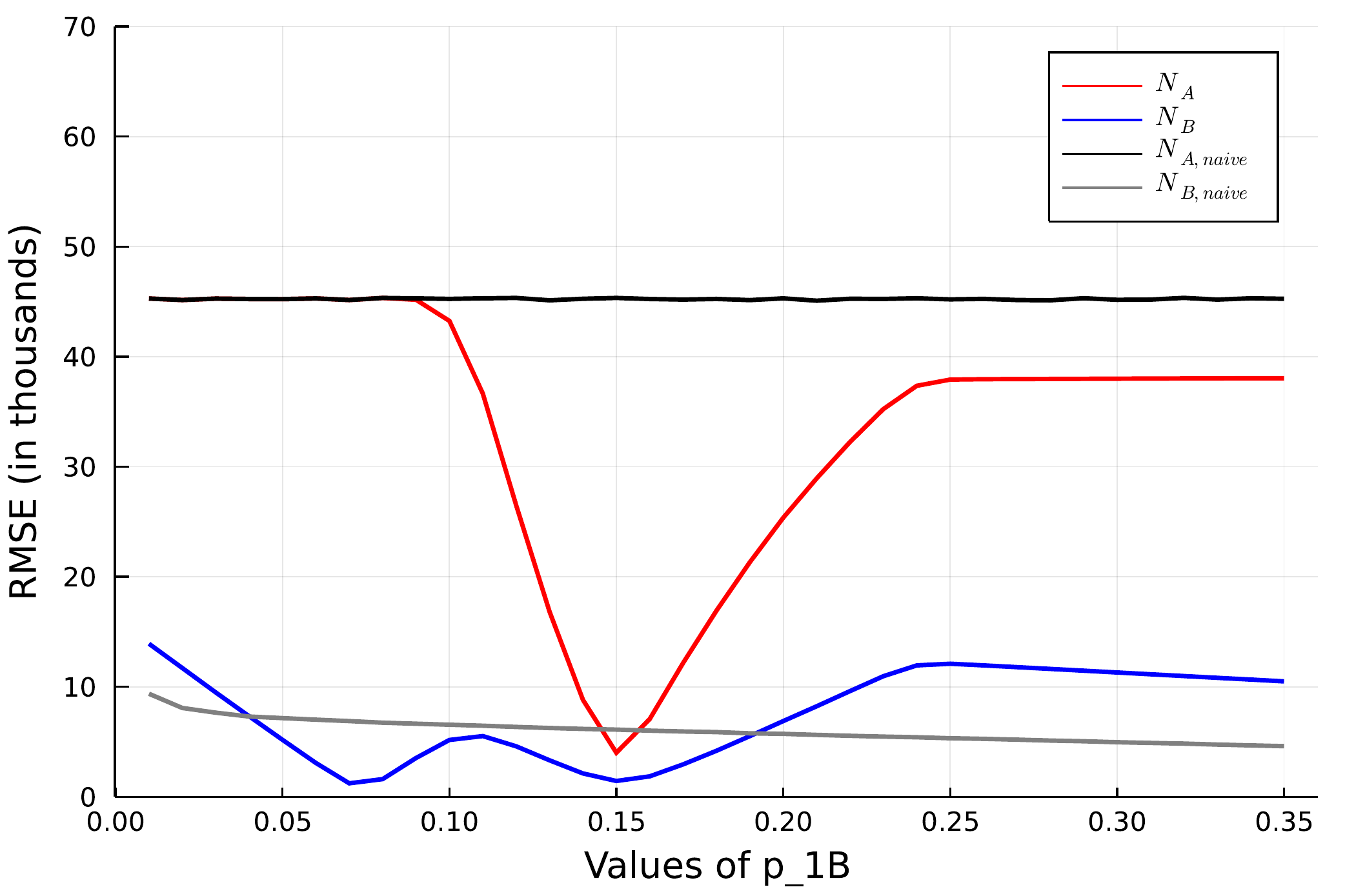}
    \includegraphics[width=0.45\textwidth]{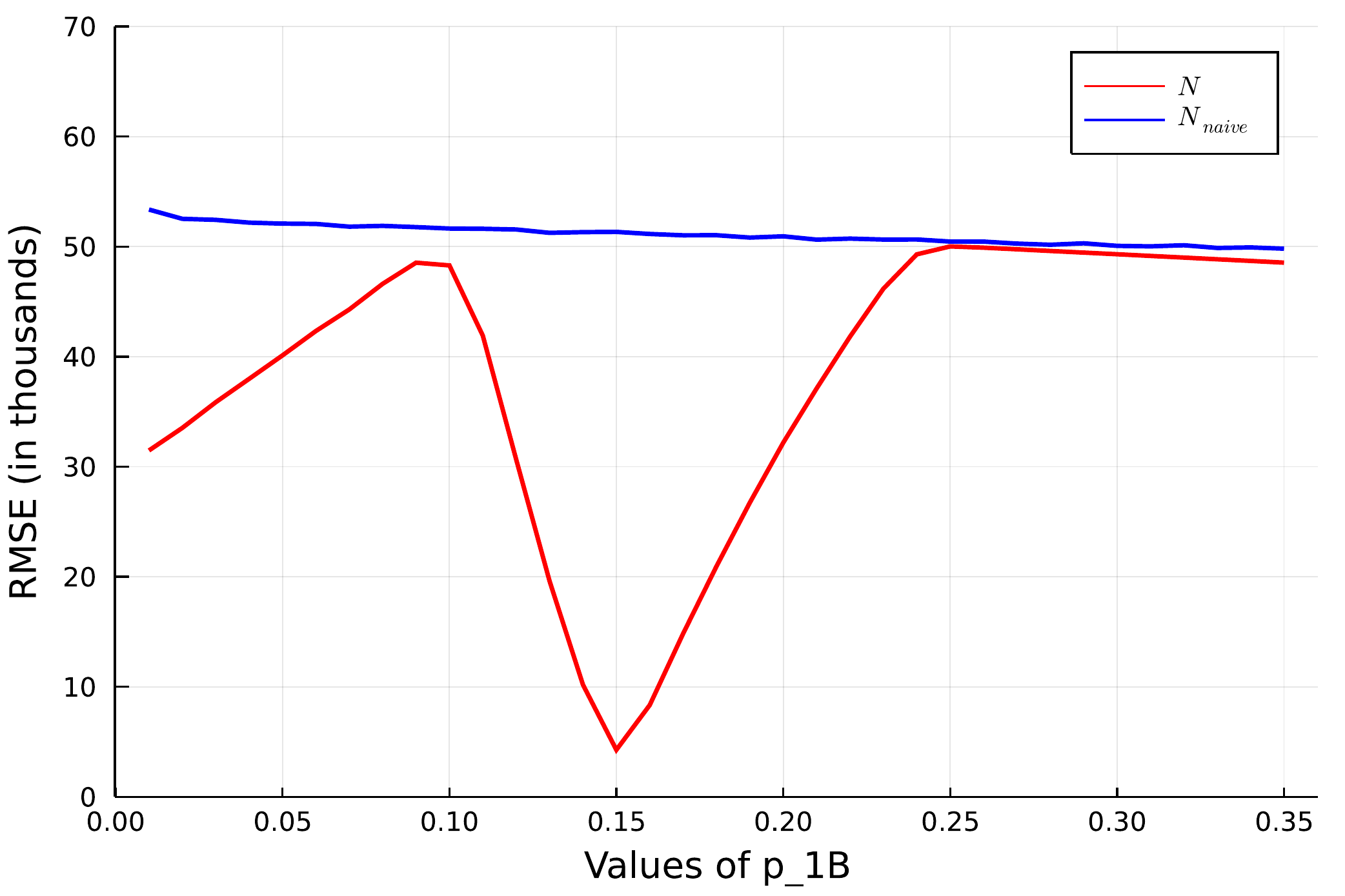}
    \caption{Bias (top) and RMSE (bottom) of the proposed estimator depending on the value of $p_{1B}$ by sub-populations (left) and the whole population (right) based on scenario 2}
    \label{appen-fig-sim2-s2}
\end{figure}

\begin{figure}[ht!]
    \centering
    \includegraphics[width=0.45\textwidth]{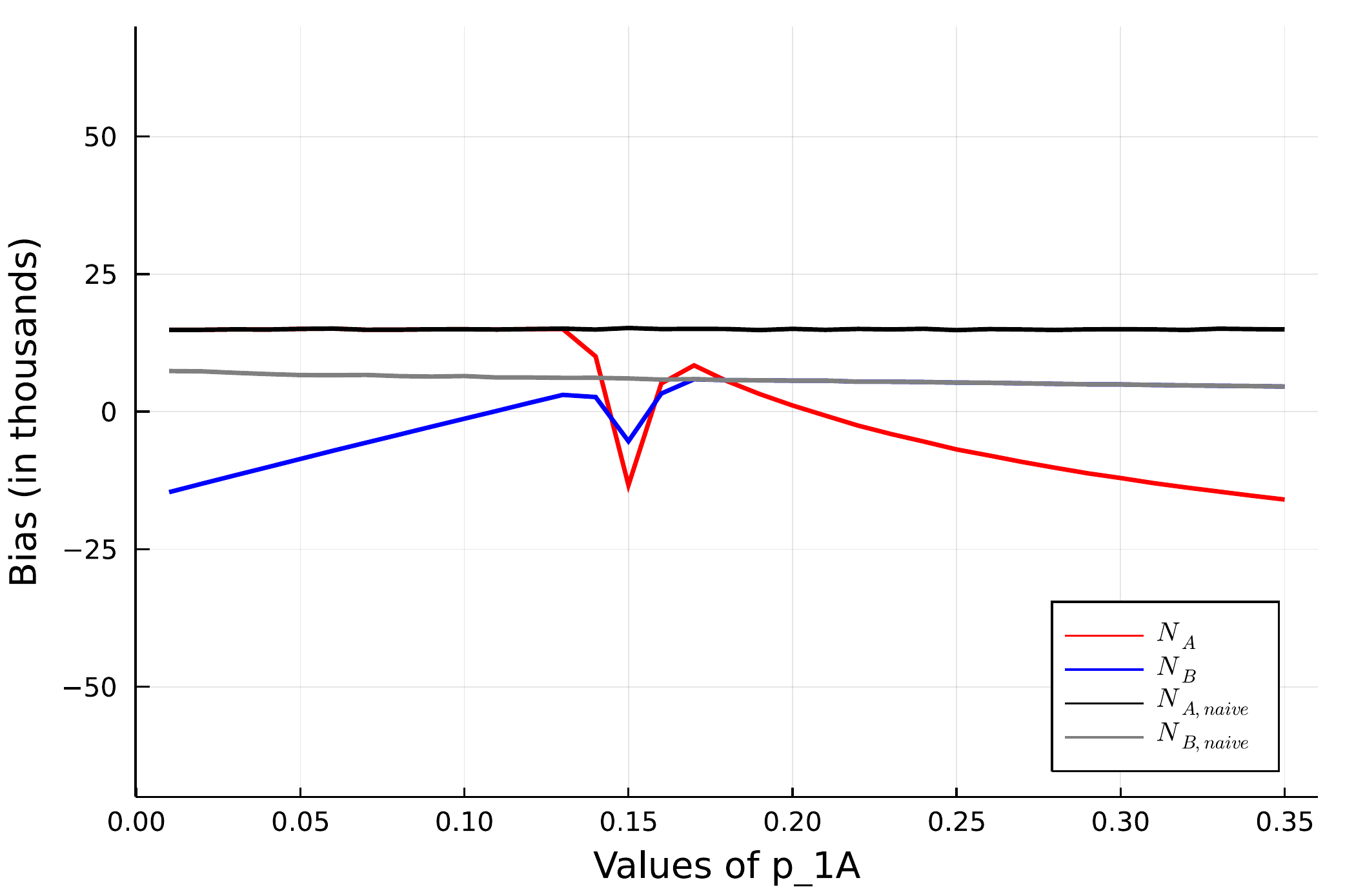}
    \includegraphics[width=0.45\textwidth]{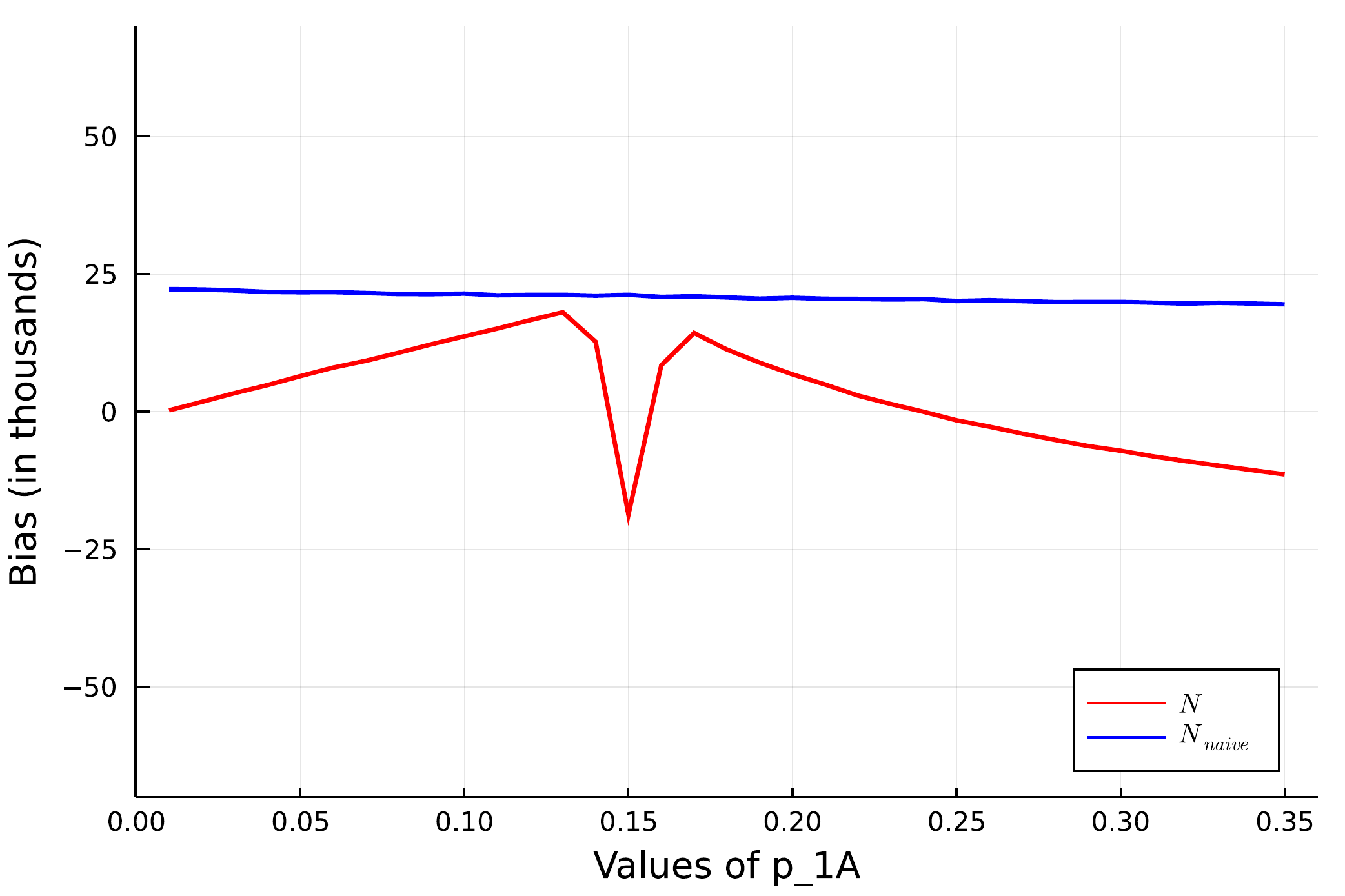}
    \includegraphics[width=0.45\textwidth]{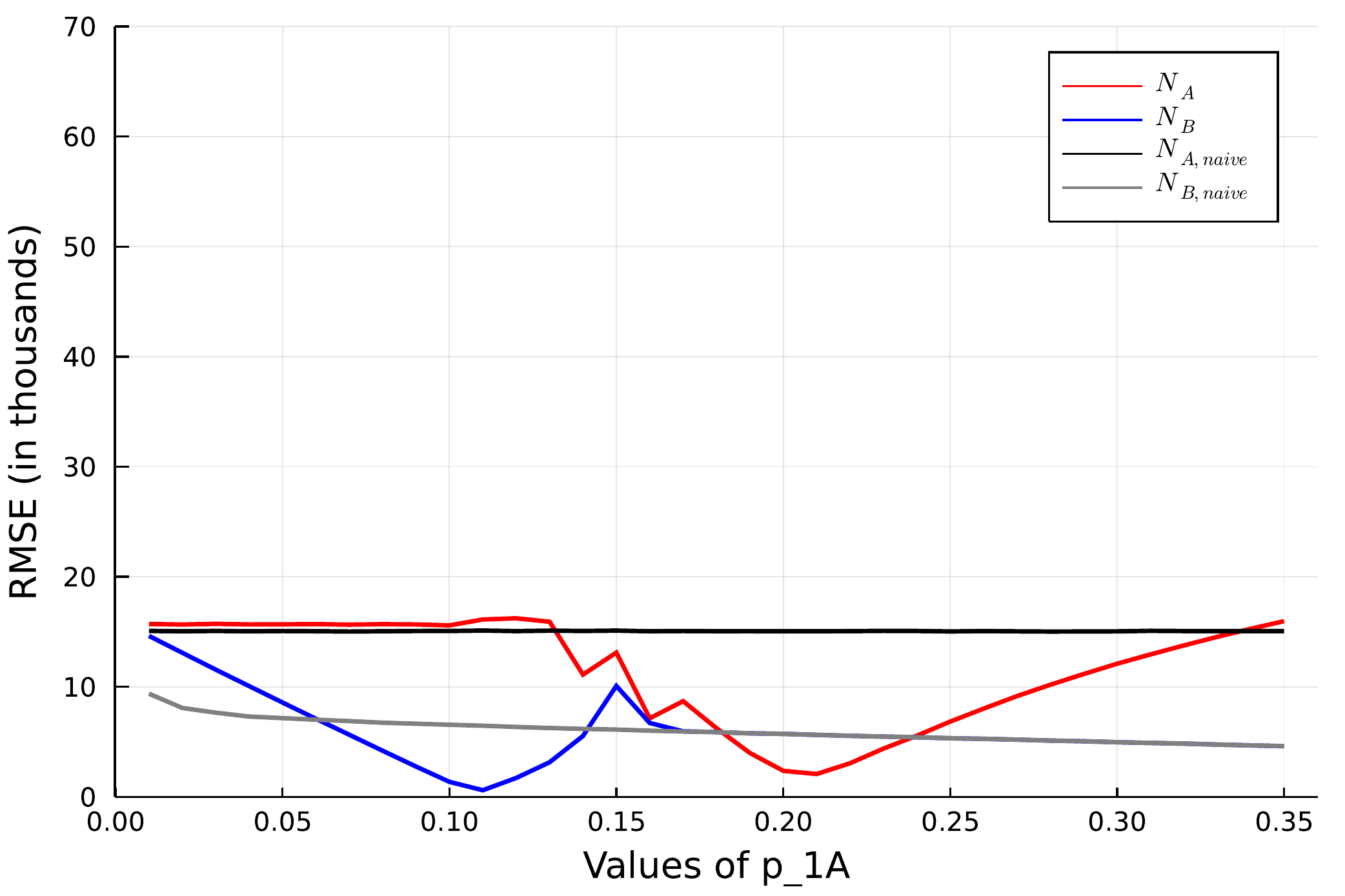}
    \includegraphics[width=0.45\textwidth]{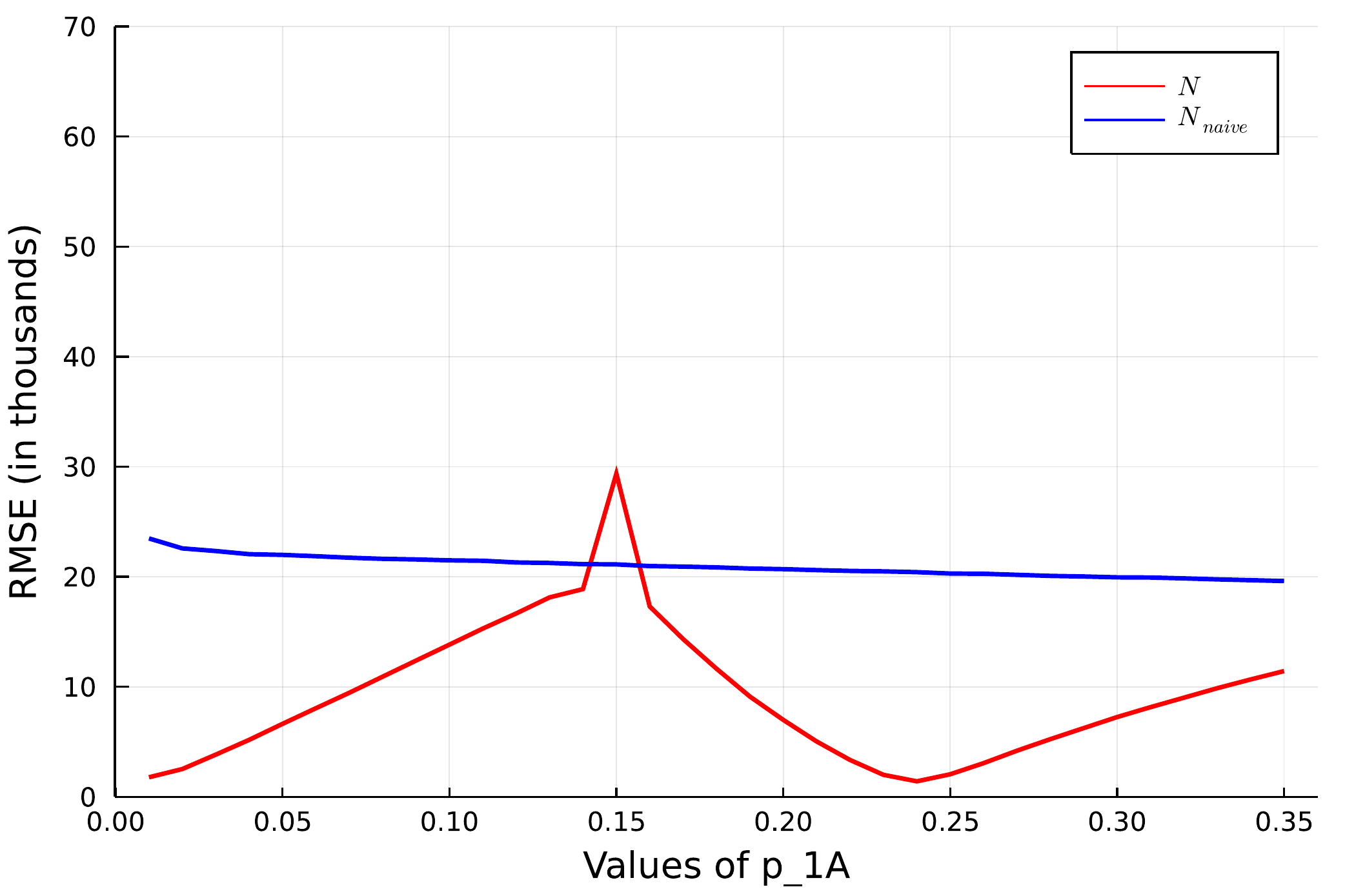}
    \caption{Bias (top) and RMSE (bottom) of the proposed estimator depending on the value of $p_{1A}$ by sub-populations (left) and the whole population (right) based on scenario 3}
    \label{appen-fig-sim2-s3}
\end{figure}

Figures \ref{appen-fig-sim2-s1} and  \ref{appen-fig-sim2-s2} indicate that bias increases with the growing difference between $p_{1A}$ and $p_{1B}$. If the value of $p_{1A},p_{1B}$ is closer to the nominal value of 0.15, bias disappears. On the other hand, when $p_{1A}=p_{1B}=0.15$, bias is present even if $p_{1A}=p_{1B}$ (this results holds also for other values of $p_{1A}=p_{1B}$)) as presented in Figure \ref{appen-fig-sim2-s3}. Thus, the estimator is not robust when the main assumption is not met, but as we use constrained MLE, bias is lower than the one of the nai\"ve estimator.

\end{document}